\documentclass[aps,pra,twocolumn,superscriptaddress,a4paper,notitlepage]{revtex4-1}

\usepackage{amssymb,algpseudocode,algorithm}
\usepackage{amsmath}
 \usepackage{color}
\usepackage{psfrag}
\usepackage{ifpdf}
\ifpdf
\usepackage{epstopdf}   
\fi

 \newcommand{\ket}[1]{\ensuremath{\vert#1\rangle}}
 \newcommand{\bra}[1]{\ensuremath{\langle #1\vert}}
\newcommand{\bk}[2]{\ensuremath{\langle #1\vert #2\rangle}}
\newcommand{\kb}[2]{\ensuremath{\vert #1 \rangle \langle #2 \vert}}

\newcommand{\ptr}{\ensuremath{\mathrm{tr}_a}}

\def\id{{\mathchoice {\rm 1\mskip-4mu l} {\rm 1\mskip-4mu l} {\rm 1\mskip-4.5mu l} {\rm 1\mskip-5mu l}}}

\begin{document}

\title{An efficient magic state approach to small angle rotations}

\author{Earl T. Campbell}
 \affiliation{Department of Physics \& Astronomy, University of Sheffield, Sheffield, S3 7RH, United Kingdom.}
 \email{earltcampbell@gmail.com}
\author{Joe O'Gorman}
 \affiliation{Department of Materials, University of Oxford, Oxford, OX1 3PH, United Kingdom.}
 
\begin{abstract}
Standard error correction techniques only provide a quantum memory and need extra gadgets to perform computation. Central to quantum algorithms are small angle rotations, which can be fault-tolerantly implemented given a supply of an unconventional species of magic state.  We present a low-cost distillation routine for preparing these small angle magic states.  Our protocol builds on the work of Duclos-Cianci and Poulin [\textit{Phys. Rev. A} \textbf{91}, 042315 (2015)] by compressing their circuit. Additionally, we present a method of diluting magic states that reduces costs associated with very small angle rotations.  We quantify performance by the expected number of noisy magic states consumed per rotation, and compare with other protocols. For modest size angles, our protocols offer a factor 24 improvement over the best known gate synthesis protocols and a factor 2 over the Duclos-Cianci and Poulin protocol.  For very small angle rotations, the dilution protocol dramatically reduces costs, giving several orders magnitude improvement over competitors.  There also exists an intermediary regime of small, but not very small, angles where our approach gives a marginal improvement over gate synthesis.  We discuss how different performance metrics may alter these conclusions. 
\end{abstract}

\maketitle 

\section{Introduction}

Fault-tolerant quantum computing involves a host of resource overheads, entering at different stages of the process.  The most widely known cost is that of encoding a logical qubit into many physical qubits, which provides safe storage of quantum information.  However, once encoded, a logical qubit only natively supports a limited set of fault-tolerant operations~\cite{Eastin09}.  For high-threshold codes, the native operations belong within the Clifford group.  Additional layers of gadgets are required to enable general purpose quantum computation.  A  corollary of the Solovay-Kitaev theorem~\cite{kitaev02} is that if the Clifford group is augmented by a non-Clifford $T$ gate, also called the $\pi/8$ gate, the device can efficiently approximate any required quantum algorithm. These $T$ gates can in turn be fault-tolerantly performed by using state injection of high-fidelity magic states, prepared by distillation~\cite{BraKit05}.  However, the Solovay-Kitaev theorem requires a huge number of $T$ gates, and early magic state distillation protocols put a high price on each one.  The last decade has seen substantial advances in both these areas.  Magic state distillation is now possible at better rates~\cite{BraKit05,Meier13,Bravyi12,Jones13,fowler13}, reducing the expected cost per high quality $T$-gate.  The number of $T$-gates needed to approximate some unitary, the so-called $T$-count, has also been significantly reduced after the discovery of new gate synthesis techniques~\cite{kliuchnikov13,paetznick14,gosset14,RS14,amy16}.

\begin{figure}[t]
    \includegraphics{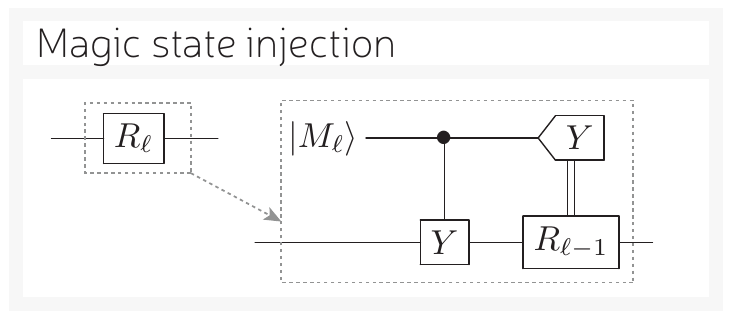}
    \caption{State-injection using a magic state $\ket{M_\ell}$ to perform a non-Clifford gate $R_\ell$.  If the $Y$-basis measurement outcome is ``+1", then a correction $R_{\ell-1}$ is needed.}
    \label{FIG_inject}
\end{figure}

Nevertheless, fault-tolerant quantum computing remains a monumental challenge, so further resources savings are essential.  One suggested route is to circumvent magic states altogether, for instance by using gauge-fixing~\cite{Anderson14,bombin13b} of subsystem codes like the 3D colour code~\cite{bombin06,Bombin13}.  However, this route still pays a price for gate-synthesis and the additional qubit cost of 3D codes leads to a cubic scaling of overheads that is also achievable using magic states~\cite{RHG01a,Gorman12}. Furthermore, all current indications are that colour codes have a much lower error correction threshold~\cite{brown15} than the toric code~\cite{wang03,Rauss07,RHG01a,fowler12b}.  For now, such low-noise levels appear beyond technological reach, ruling out gauge-fixing in the near-term.  Alternatively, one could work with qudits, $d$-level systems, which are favourable in terms of the cost of magic state distillation~\cite{Anwar12,Campbell12,campbell14,dawkins15}, though in the qudit context little is known about gate synthesis or experimental feasibility.  

Landahl and Cesare~\cite{landahl13} were the first to suggest that the 
gate synthesis overhead could be reduced by distilling different species of magic states, which provide smaller angle rotations instead of $T$-gates.  Specifically, they considered  $\exp (i \pi Z /2^\ell )$ gates for integer $\ell$, where $\ell=3$ gives the $T$-gate.  These gates are part of an important family called the Clifford hierarchy~\cite{gottesman99}, with $\exp (i \pi Z /2^\ell )$ residing in the $\ell^\mathrm{th}$ level of the hierarchy and naturally appearing in quantum simulation~\cite{lloyd96,poulin14,trout15,wecker15,hastings15} and the quantum Fourier transform. Landahl and Cesare found that for small integer $\ell$, distillation was favourable over gate synthesis methods known at the time.  Unfortunately, for very small angle rotations $5 \lesssim \ell $, they saw this advantage vanish.     

\begin{figure*}
    \includegraphics{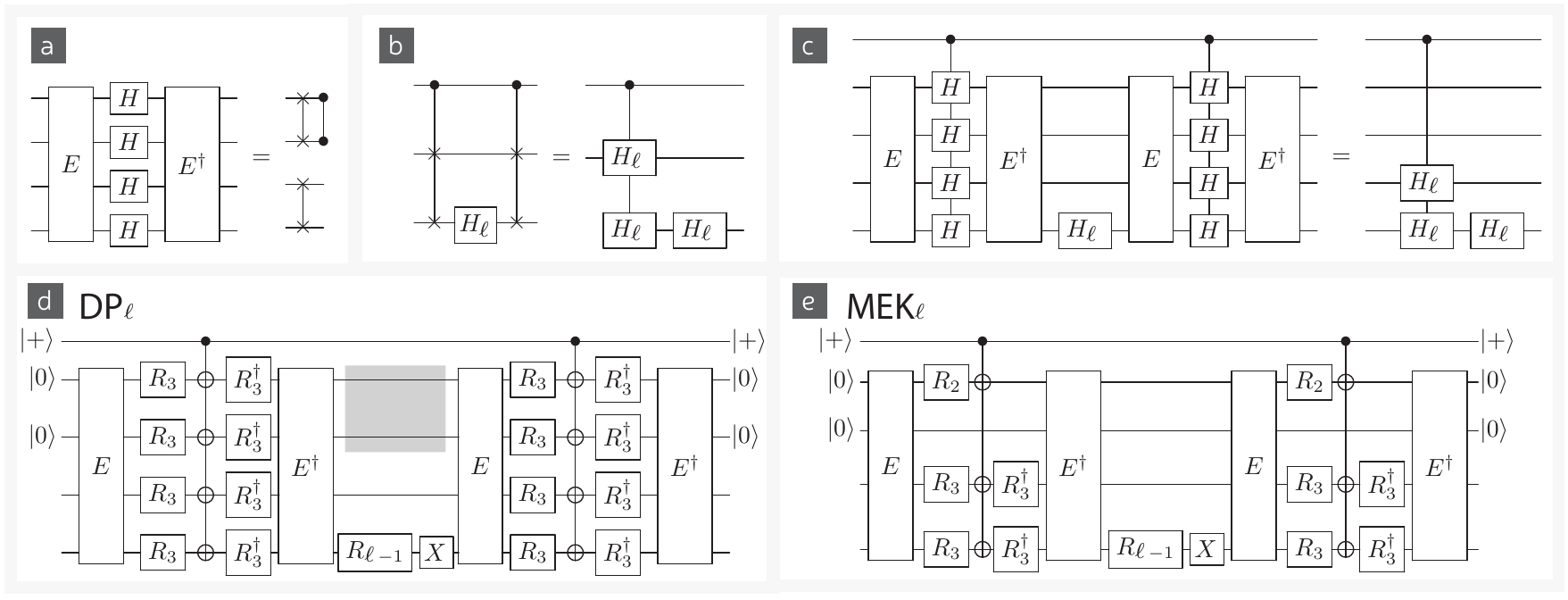}
    \caption{(a,b,c) Circuit identities and gadgets used to construct distillation protocols. (d) a non-compressed distillation circuit. Adding $\kb{0}{0}$ measurements and preparations within the grey box gives the DP$_\ell$ protocol. (e) the compressed distillation circuit describing MEK$_\ell$.  In both (d) and (e) noisy $\ket{M_\ell}$ states are input on the bottom two circuits lines (labelled qubits 3 and 4 in the main text) and all $R_\ell$ gates with $L\geq 3$ are approximately implemented by injection of a noisy $\ket{M_\ell}$ states. }
    \label{FIG_IDENTs1}
\end{figure*}

Progress in gate synthesis has significantly eroded the importance of the Landahl and Cesare result, though their general idea endures.  Several researchers found efficient protocols that distill Toffoli magic states~\cite{jones13b,eastin13,Paetznick13}.  Duclos-Cianci and Poulin~\cite{duclos15} modified a protocol proposed by Meier, Eastin and Knill~\cite{Meier13} to prepare magic states that again provide small angle rotations.  This recent result was a substantial leap forward in the art of distilling magic states for small angle rotations, making magic state distillation cheaper than gate synthesis in many instances.  We revisit the work of Duclos-Cianci and Poulin, and find their protocol can be further refined.  Our proposed alternative requires fewer resources per attempt, and achieves superior error suppression with a higher success probability.  Both our protocol and that of Duclos-Cianci and Poulin build on the earlier work of Meier, Eastin, and Knill (MEK)~\cite{Meier13}. 

For very small angle rotations, the required magic states become close to stabiliser states.  Duclos-Cianci and Poulin observed that in this regime some magic states in the protocol can be supplanted by stabiliser states, reducing resource costs in a certain regime.  Instead, we introduce the notion of magic state dilution, which takes magic states for $\ell$ level rotations, and converts them into a larger number of magic states for a finer rotations  with higher $\ell$. Error rates are adjusted in the dilution process, and we find it works best at high $\ell$.  Remarkably, dilution can even reduce noise when used at sufficiently high $\ell$.

\section{Magic states model}

First we review the Clifford hierarchy and the magic state model.  The well known Pauli group $\mathcal{P}$ is the group composed of tensor products of the single qubit Pauli operators.  Unitaries mapping the Pauli group to itself are called Clifford unitaries, which again form a group $\mathcal{C}:= \{ U | U P U^\dagger \in \mathcal{P}, \forall P \in \mathcal{P} \}$.  Clifford operations are  physical operations composable from Clifford unitaries, measurement of Pauli operators and preparation of their eigenstates (the stabiliser states), and classical feedforward.  The magic state model of quantum computation~\cite{BraKit05} assumes Clifford operations are free resources that can be implemented perfectly, and proceeds to evaluate the cost of non-Clifford operations.  Such a model is suitable for logical qubits where the Clifford operations are fault-tolerantly protected, as is common.  In our conclusions, we discuss further the validity of counting only non-Clifford resources.

Unitaries outside the Clifford group can fall into other levels of the Clifford hierarchy, defined recursively as
\begin{equation}
    \mathcal{C}_\ell := \{ U | U P U^\dagger \in \mathcal{P}, \forall P \in \mathcal{C}_{\ell-1} \},
\end{equation}
where $\mathcal{C}_1 := \mathcal{P}$. It follows that the Clifford group is the second level, and all higher levels include non-Clifford gates.  We will soon see the Clifford hierarchy plays an important operational role in a teleportation process known as state injection~\cite{GC01a}.

We introduced small angle rotations in the $Z$ basis, but here it is more convenient to work in the $Y$ basis. We define unitaries $R_\ell := \exp (i \theta_\ell Y )$ with $\theta_\ell = \pi/2^\ell$.  The $R_1$ and $R_2$ gates are Clifford, which are presumed ideal and an inexpensive resource.  The $R_3$ gate is non-Clifford and equivalent to the $T$ gate.  For $\ell \geq 3$, the gates are non-Clifford and belong to the $\ell^{\mathrm{th}}$ level of the Clifford hierarchy.  Also important here are certain Hermitian operators in the Clifford hierarchy, and we define  $H_\ell := R_\ell X R_\ell^\dagger= R_\ell^2 X  =R_{\ell-1} X  $ and find $H_\ell \in \mathcal{C}_{\ell-1}$. We remark that $H_3$ equals the Hadamard.
 
Preparation of non-stabiliser states is also a non-Clifford operation, and these magic states also fall into a natural hierarchy.  Recall that stabiliser states are eigenstates of Pauli operators, which are elements of $\mathcal{C}_{1}$. For example, the computational basis sates $\ket{0}$ and $\ket{1}$ are stabilised by the $Z$ and $-Z$ Pauli operators, respectively.  Below we will also make use of the stabiliser states $\ket{\pm} = ( \ket{0} \pm \ket{1}) / \sqrt{2} $, which satisfy $(\pm X) \ket{\pm} = \ket{\pm}$ for Pauli $X$. We consider magic states that are eigenstates of the $H_\ell$ operators, so that
\begin{align}
    \ket{M_\ell} &=  H_\ell \ket{M_\ell} =  R_\ell \ket{+} , \\ 
    \ket{\bar{M}_\ell} &=  (-H_\ell) \ket{\bar{M}_\ell} =  R_\ell \ket{-}, 
\end{align}
which we refer to as $\ell^{\mathrm{th}}$ level magic state states.  Such resources can be used to inject $R_\ell$ rotations into circuits as shown in Fig.~(\ref{FIG_inject}).  The injection procedure is probabilistic, and with probability $1/2$ it performs $R_\ell$ and with probability $1/2$ it performs $R_\ell^\dagger$.  In the latter case a correction of $R_\ell^2 = R_{\ell-1}$ is needed to get the desired result. Since $R_2$ is Clifford, the injection process is ensured to terminate within $\ell-2$ attempts.  

Therefore, one can accomplish small angle rotations with a cost that depends on the cost of distilling high-fidelity $\ket{M_\ell}$ states.  In contrast, gate synthesis methods prescribe distilling just $\ket{M_3}$ states and finding a gate sequence $R_\ell  \approx C_1 T C_2 \ldots T C_n$ in terms of $T$ gates and Cliffords $C_j$.  Although, $\ket{M_\ell}$ states will prove more expensive than $\ket{M_3}$, gate synthesis can require very many $\ket{M_3}$ states, making it more expensive overall.  There are many gate synthesis algorithms for finding gate sequences and here we consider the Selinger and Ross~\cite{RS14} algorithm (SR) and probabilistic quantum circuits with fallback (PQF)~\cite{bocharov15}.  SR has the benefit of a freely available software implementation and is provable optimal for unitary synthesis.   PQF makes use of ancilla and measurements to achieved the best known performance, about a factor 3 more efficient than SR.  Several other methods (see e.g.~\cite{paetznick14,bocharov15b}) exist, but all fall somewhere between SR and PQF in performance.  We discuss the applicability of our results to rotations of arbitrarily angles in App.~\ref{Arb_angle}.

\section{Overview of resource costs}

Next we summarise the work of Meier, Eastin, and Knill (MEK)~\cite{Meier13}, upon which Duclos-Cianci and Poulin built their construction.  The MEK protocol takes 10 input magic states and with some probability outputs 2 magic states.  Therefore, MEK is said to be a $10 \rightarrow 2$ protocol.    Accounting for the species of magic states, we call MEK a $10_3 \rightarrow 2_3$ protocol where the subscripts indicate that MEK both inputs and outputs 3rd level magic states.  We use DP$_{\ell}$ to label the Duclos-Cianci and Poulin protocol for distilling $\ell^\mathrm{th}$ level magic states.  Each round of DP$_{\ell}$ consumes a cocktail of different input magic states, and we describe it as a
\begin{equation}
\left\{ 16_3, 2_\ell, 1_{\ell-1}, \left(\frac{1}{2}\right)_{\ell-2},  \left(\frac{1}{4}\right)_{\ell-3}, \ldots \left(\frac{1}{2^{\ell-3}}\right)_{3} \right\} \rightarrow 2_\ell
\end{equation} 
protocol.  Again, subscripts show what level magic state is used.  We show the expected number of inputs required per attempt, which is sometimes a fraction.  The fractional sequence terminates at the third level, because lower levels correspond to Clifford resources and are considered free.  The DP$_{\ell}$ protocol is presented as a direct generalisation of MEK, but one sees that in the case of $\ell=3$ the DP$_{3}$ protocol is a $18_3 \rightarrow 2_3$ protocol and so actually needs almost twice as many input states as MEK.   Here we construct a streamlined variant of DP$_\ell$, that we call MEK$_\ell$ and is a 
\begin{equation}
 \left\{ 8_3, 2_\ell, 1_{\ell-1},  \left(\frac{1}{2}_{\ell-2}\right),  \left(\frac{1}{4}\right)_{\ell-3}, \ldots \left(\frac{1}{2^{\ell-3}}\right)_{3}  \right\} \rightarrow 2_\ell
\end{equation} 
protocol.  We find that MEK$_3$ corresponds precisely to the original MEK, so that MEK$_\ell$ is a more apt generalisation than DP$_{\ell}$.  Not only does MEK$_\ell$ require fewer input resources, it also achieves superior error suppression with a higher success probability.

Both MEK and DP$_{\ell}$ make use of a simple 4 qubit code.  We use $E$, short for encoder, to denote the Clifford circuit that brings qubits into the codespace and acts on pairs of Pauli operators as
\begin{align}
\label{Cliff}
    (Z_1 , X_1 ) & \rightarrow ( Z_1 Z_2 Z_3 Z_4, X_1 X_3 X_4 )  ,\\
    (Z_2 , X_2 ) & \rightarrow ( X_1 X_2 X_3 X_4  , Z_2 ) , \\
    (Z_3 , X_3 ) & \rightarrow ( Z_1 Z_4 , X_1 X_3 ), \\
    (Z_4 , X_4 ) & \rightarrow ( X_1 X_4 , Z_1 Z_3 ).
\end{align}
For Pauli operators, we always use the subscript to denote which qubit the operator acts upon. Preparing the first two qubits in the state $\ket{0}$ and running the encoder will yield the code stabilised by $Z_1 Z_2 Z_3 Z_4$ and $X_1 X_2 X_3 X_4$.  The logical state in the codespace is determined by the initial states of the last two qubits.  Having defined the codespace, we now turn to describing DP$_{\ell}$ in more detail, where we will also discuss the important features of the 4-qubit code.

\begin{figure*}
    \includegraphics{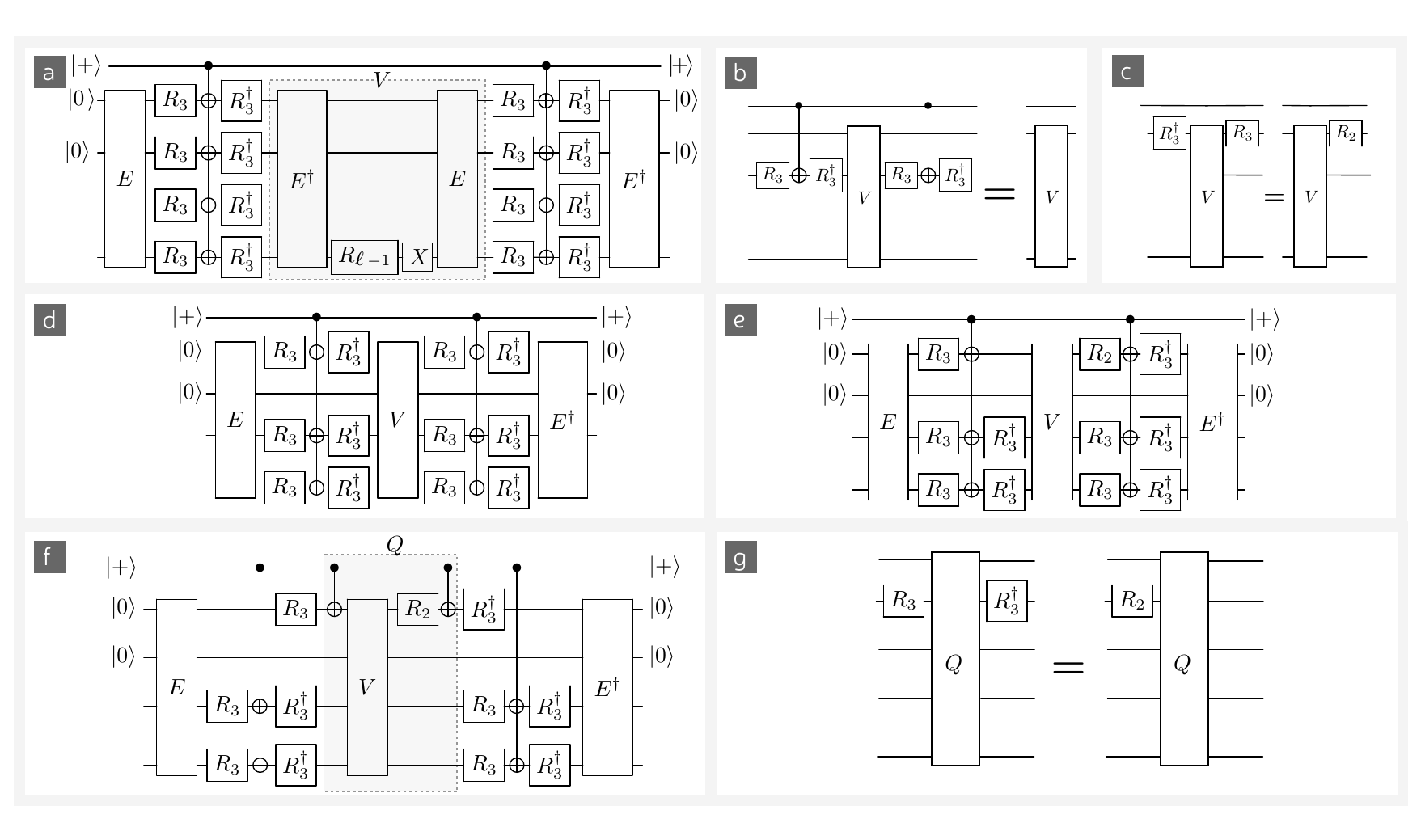}
    \caption{Circuit identities and reductions used to obtain MEK$_\ell$. (a) circuit before compression, with subcircuit $V$ labelled.  (b) and (c) show properties of $V$. (d) shows partially compressed circuit after applying identity (b).  (e) further compressed circuit after applying identity (c).  In (f) we identify subcircuit $Q$, with (g) showing a property of $Q$.  Applying (g) to (f) yields the final compressed circuit, shown in Fig.~(\ref{FIG_IDENTs1}e).}
    \label{FIG_compress}
\end{figure*}

\section{Uncompressed DP$_{\ell}$ protocol}

The protocol DP$_{\ell}$ is illustrated in Fig.~(\ref{FIG_IDENTs1}d). Throughout, we label the top wire as the control qubit $c$, and the subsequent qubits are labelled 1 to 4.  We summarise the main steps of the protocol here
\begin{enumerate}
    \item Prepare qubits $c$, 1 and 2, in stabiliser states $\ket{+}$, $\ket{0}$ and $\ket{0}$, respectively;
    \item Prepare qubits 3 and 4 in noisy $\ket{M_\ell}$ states with $\epsilon_\ell$ error;
    \item Perform the circuit gates shown in Fig.~(\ref{FIG_IDENTs1}d): the $R_3$ gates are achieved by state injection using eight $\ket{M_3}$ states, and $R_{\ell-1}$ is achieved with state injection resulting in overall $\eta_{\ell-1}$ error;    
    \item Qubit $c$ is measured in $X$ basis, and we continue if the outcome is $``+1"$ and otherwise declare FAILURE;
    \item Qubits 1 and 2 are measured in $Z$ basis, and we declare SUCCESS if the outcome is $``+1"$ and otherwise declare FAILURE;
    \item If successful, qubits 3 and 4 are output as $\ket{M_\ell}$ states of higher fidelity.\end{enumerate}
Note that step 3 can include some additional postselected measurements  highlighted in Fig.~(\ref{FIG_IDENTs1}d).  If included these measurements detect some additional errors, but even without these measurements the protocol quadratically suppresses noise. For simplicity we will assume they are not performed. We next review some of the basic intuition behind why DP$_{\ell}$ works, which is closely related to properties of the 4 qubit code used. 

The codespace provides both important transversality features and keeps the protocol protected against certain faults.  Regarding transversality, a global Hadamard $H \otimes H \otimes H \otimes H $ will preserve the code space, implementing a logical SWAP between the two encoded logical bits.  More generally, one can verify that $E ( H \otimes H \otimes H \otimes H ) E^\dagger$ acts as shown in Fig.~(\ref{FIG_IDENTs1}a). We see that even without fixing the first two qubits to $\ket{0}$, this circuit implements a swap and a phase-swap (a swap combined with a phase gate).  Furthermore, implementing controlled Hadamards within the encoding will realise controlled versions of the swap and phase-swap. Next, we note that for any Hermitian unitary, such as $H_\ell$, we have that conjugation by controlled-swaps will produce a controlled $H_\ell \otimes H_\ell$ as shown in Fig.~(\ref{FIG_IDENTs1}b).  We combine this observation with the transversality properties of the code to get the identity of Fig.~(\ref{FIG_IDENTs1}c).  An ancilla on the control of the controlled $H_\ell \otimes H_\ell$ is used to measure the $H_\ell \otimes H_\ell$ observable.  If the desired magic state is an eigenstate of $H_\ell$, then measuring $H_\ell \otimes H_\ell$ allows us to detect a single error between two noisy $\ket{M_\ell}$ states input on the bottom two circuit lines.  In this sense, the codespace has provided a SWAP gadget for distilling noisy $\ket{M_\ell}$  magic states.

Although the controlled-Hadamard rotations and $H_\ell$ rotation may be non-Clifford, noisy magic states can be used to implement these operations.  To see this recall that $H= R_3 X R_3^\dagger$, and so a control-Hadamard can be implemented by a control-$X$ gate sandwiched between $R_3$ and $R_3^\dagger$.  In turn, $R_3$ and $R_3^\dagger$ can each be implemented at the cost of a single $\ket{M_3}$ magic state.  Given noisy $\ket{M_3}$ magic states, we implement noisy $R_3$ gates.   This brings us to the second role played by the error-correction code. The $R_3$ gates are performed on qubits within an error correction code that can detect a single qubit error, and so we will detect a single error in any $\ket{M_3}$ magic states. Let us explain this point in more detail.  When using state injection, if a magic state carries an error, then it will result in $Y \cdot R_3$ instead of $R_3$, and so there is an additional $Y$ acting on one of the four qubits.  Inside the encoding, the state is an eigenstate of $X\otimes X \otimes X \otimes X$,  but a $Y$ error will cause the state to become an eigenstate of $-X\otimes X \otimes X \otimes X$.  At the end of the circuit, we decode and measure, which is equivalent to measuring the $X\otimes X \otimes X \otimes X$ observable.  Since we postselect on all $+1$ outcomes,  any error on a single $\ket{M_3}$ will be detected. In contrast, the central $R_{\ell-1}$ rotation occurs outside the protection of the codespace when the logical qubits have been decoded onto single qubits.  Therefore, the protocol will not be robust against failure of this gate, and so this rotation must be high fidelity and we herein call it the pivotal rotation.  Nevertheless, we can construct good distillation protocols provided magic states used in the pivotal rotation have already been distilled to a much higher fidelity than all other elements of the circuit.  These high fidelity resources will be costly, but the protocol remains efficient because resources for performing $R_{\ell-1}$ are less valuable than the $\ket{M_\ell}$ states we are distilling.  We are erecting a pyramid of distillation protocols, with resources from lower in the Clifford hierarchy fuelling distillation at higher levels.

\section{Compressed MEK$_{\ell}$ protocol}

Our main contribution is to show that this circuit can be compressed into Fig.~(\ref{FIG_IDENTs1}e). This cancels several $R_3$ rotations to reduce the number of magic states consumed. The steps of the protocol roughly follows those numerically listed in the previous section, except step 3 now uses the circuit of Fig.~(\ref{FIG_IDENTs1}e), and we use only 8 magic states to inject the $R_3$ gates.   It is important that the circuit retains its fault-tolerance properties through the compression process. That is, even when compressed the noisy $R_3$ gates still occur within the four-qubit error correction code, and so we still expect to detect the failure of any single $R_3$ gate.  In App.~\ref{APP_noise}, we present a full noise analysis that rigorously confirms this intuition.  

Here we show how to compress the distillation protocol, removing unnecessary non-Clifford gates.  We start with the protocol shown in Fig.~(\ref{FIG_IDENTs1}d) and through a series of circuit reductions arrive at Fig.~(\ref{FIG_IDENTs1}e).   Taking Fig.~(\ref{FIG_IDENTs1}d), we identify a subcircuit $V$ shown in Fig.~(\ref{FIG_compress}a) inside a shaded box. Next we establish two properties of $V$, which are illustrated in Figs.~(\ref{FIG_compress}b) and~(\ref{FIG_compress}c). 
Algebraically, the $V$ circuit is simply $V=E \exp(i \theta_{\ell-1} Y_4)X_4 E^{\dagger}$, and using Eqs.~(\ref{Cliff}) we see that $V=\exp(-i \theta_{\ell-1} Y_1 Z_3 X_4) Z_1 Z_3 $.  This acts trivially on the second qubit and control qubit, entailing the circuit identity of Fig.~(\ref{FIG_compress}b).  Going further, we notice that $V Y_1 = - Y_1 V$ and so $\exp(i \theta_3 Y_1) V \exp(- i \theta_3 Y_1) = \exp(i 2 \theta_3 Y_1)  V $.  Using $2 \theta_3 = \theta_2 $, we conclude  $\exp(i \theta_3 Y_1) V \exp(- i \theta_3 Y_1) =   \exp(i 2 \theta_2 Y_1) V$.  Recall that $\exp(i 2 \theta_2 Y_1)$ is the $R_2$ gate acting on qubit 1, and so we have the identity shown in Fig.~(\ref{FIG_compress}c).

Applying the identity Fig.~(\ref{FIG_compress}b), shows that Fig.~(\ref{FIG_compress}d) is equivalent to the original circuit.  This observation has eliminated 4 non-Clifford gates.  Next, we apply the identity of Fig.~(\ref{FIG_compress}c), to obtain Fig.~(\ref{FIG_compress}e). Since $R_2$ is Clifford, we have removed 2 further non-Cliffords from the circuit.

Next, we group together a new collection of circuit elements $Q$ shown in dashed box of Fig.~(\ref{FIG_compress}f). Algebraically, $Q$ is
\begin{align}
Q &= C^{X}_{c,1} \exp(i  \theta_2 Y_1 )  V  C^{X}_{c,1}  \\ \nonumber
&= C^{X}_{c,1} \exp(i  \theta_2 Y_1 ) \exp(-i \theta_{\ell-1} Y_1 Z_3 X_4) Z_1 Z_3  C^{X}_{c,1} \\ \nonumber
 &= \exp(i  \theta_2 Z_c Y_1  )  \exp(-i \theta_{\ell-1} Z_c Y_1  Z_3 X_4)Z_c Z_1  Z_3  
\end{align}
where we have used $C^{X}$ for control-$X$ gates and their conjugation relations $C^{X}_{c,1} X_c C^{X}_{c,1}=X_c X_1 $ and $C^{X}_{c,1} Z_1 C^{X}_{c,1}=Z_c Z_1   $.  From this expression for $Q$ we can again see that $Q Y_1 = - Y_1 Q$, which entails that $\exp(- i \theta_3 Y_1) Q \exp( i \theta_3 Y_1) = Q \exp( i 2 \theta_3 Y_1) =Q \exp( i  \theta_2 Y_1)$.  This demonstrates the identity of Fig.~(\ref{FIG_compress}g).  Applying this identity to Fig.~(\ref{FIG_compress}f), yields the final representation of MEK$_\ell$ as shown in Fig.~(\ref{FIG_IDENTs1}e).  

\section{Measuring performance}

This section introduces several definitions and notations used to describe the performance of protocols, and quantify their cost.

\subsection{Quantifying noise}
\label{quant_noise}

If $\rho$ is a noisy $\ket{\psi}$ state then we say it has error rate $\epsilon$ where $\epsilon :=\frac{1}{2} || \rho - \kb{\psi}{\psi} ||_1$ and $|| X ||_1 = \mathrm{tr}[ \sqrt{X X^\dagger} ]$ is the Schatten 1-norm.  Unless otherwise stated, we assume diagonal noise so that $\rho$ is diagonal in the same basis as $\ket{\psi}$, and for such states one can show $\epsilon=1-\bra{\psi}\rho \ket{\psi}$.  The protocols considered here use a cocktail of input resources. We use $\epsilon_\ell$ and $\epsilon_3$ to denote the input error rates of the noisy $\ket{M_\ell}$ and  $\ket{M_3}$ states used.  

Let $U$ be any rotation of the form $U=\exp(i \theta Y)$ and $\mathcal{U}$ be the associated ideal channel.  We consider this channel with $Y$ noise
\begin{equation}
\label{eqn:noisy_unitary}
	\mathcal{E}(\rho)=(1-\eta) U \rho U^\dagger + \eta Y U \rho U^\dagger Y^\dagger	,
\end{equation}
and say the channel has error rate $\eta$.  This can be easily to related to other noise metrics.  If $d$ is a unitarily invariant distance measure on channels then $d(\mathcal{U}, \mathcal{E})=d(\id, \mathcal{U}^{-1} \circ \mathcal{E})$ where $\id$ is the identity channel.  The composite channel $\mathcal{U}^{-1} \circ \mathcal{E}$ is a simple Pauli noise channel
\begin{equation}
	(\mathcal{U}^{-1} \circ \mathcal{E})(\rho)=(1-\eta)  \rho  + \eta Y  \rho  Y^\dagger	.
\end{equation}
A widely used distance measure is derived from the diamond norm~\cite{kitaev02}, so that $d_{\diamond}(\mathcal{U}, \mathcal{E}):=\frac{1}{2}|| \mathcal{U}-\mathcal{E} ||_{\diamond}$, and for Pauli noise of the above type it is well-known that $d_{\diamond}(\mathcal{U}, \mathcal{E})=\eta$.  

The diamond norm distance of channels is closely related to the 1-norm distance on states.  If a noisy magic state with $\epsilon$ error (as measured by 1-norm distance) is used to inject a rotation, then the resulting noisy rotation will have error not exceeding $\epsilon$ (as measured by the diamond norm).  Furthermore, diagonal noise on states results in diagonal noise on the rotation. Primarily, we investigate this diagonal noise, but in App.~\ref{APP_noise} show that generic noise is also suppressed by MEK$_{\ell}$.

\subsection{Distillation cost}

The output error from MEK$_\ell$ is always measured as the error on a single output qubit (ignoring correlations) and we find
\begin{align}
\label{delta_out}
    \delta_\ell(\epsilon_3, \epsilon_\ell, \eta_{\ell-1})  & \sim 8 \epsilon_3^2 + \epsilon_\ell^2 + \frac{1}{4}\eta_{\ell-1} + \ldots, \\
    P_{\mathrm{suc}}(\epsilon_3, \epsilon_\ell, \eta_{\ell-1})  & \sim 1 -8 \epsilon_3 - 2 \epsilon_\ell - \frac{1}{2}\eta_{\ell-1} + \ldots.
\end{align}
In App.~\ref{APP_EXPRESS},  we provide the full expressions as calculated by explicit simulation.  For $\eta = 0$ and $\epsilon_3=\epsilon_\ell$, these expressions are the same as those found by Meier, Eastin and Knill in their analysis of MEK.  

Within the context of a single distillation round the performance is independent of the level $\ell$, and is solely a function of the noise of the input states $\epsilon_\ell$, $\epsilon_3$ and $\eta$.  When we ask how much the input states cost, we find this can increase with $\ell$. Next, we consider many distillation rounds and combine all performance metrics into a single quantity, the expected resource cost $\mathfrak{C}(M_\ell, \delta)$ to distill a $\ket{M_\ell}$ state of $\delta$ error.  Lower $\delta$ can require more rounds of distillation, which drives up costs.  For our protocol we use that the cost is
\begin{equation}
    \mathfrak{C}(M_\ell, \delta_\ell ) =  \frac{2  \mathfrak{C}(M_\ell, \epsilon_\ell)  + 8 \mathfrak{C}(M_3, \epsilon_3 )  + \mathfrak{C}( R_{\ell-1}, \eta_{\ell-1} )}{2  P_{\mathrm{suc}}(\epsilon_3, \epsilon_\ell, \eta_{\ell-1})},
\end{equation} 
where $\delta_\ell$ obeys Eqs.~(\ref{delta_out}) and~(\ref{delta_out_exact}). In our analysis, we optimise the cost over many thousands of possible combinations of input resources by brute force search.  Notice that we capture the cost of the pivotal rotation as $\mathfrak{C}( R_\ell, \eta_\ell )$, which will in turn depend on the cost of magic states used to implement the rotation as 
\begin{equation}
    \mathfrak{C}( R_\ell, \eta_\ell ) =  \mathfrak{C}( M_\ell, \epsilon_\ell ) +  \frac{1}{2} \mathfrak{C}( R_{\ell-1}, \eta_{\ell-1} ),
\end{equation}
where 
\begin{equation}
\eta_\ell = \frac{1}{2}\epsilon_\ell + \frac{1}{2}(\epsilon_\ell(1-\eta_{\ell-1})+ (1-\epsilon_\ell)\eta_{\ell-1}).
\end{equation}
For the lowest non-Clifford levels, $\mathfrak{C}(M_3, \delta_3 )$ are found by minimising over different combinations of protocols including Bravyi-Haah~\cite{Bravyi12}, MEK$_3$~\cite{Meier13} and using the 15 qubit Reed-Muller code~\cite{BraKit05}.  Recall 
Clifford operations are considered free and perfect so that $\mathfrak{C}( M_2, 0 )=0$ and $\mathfrak{C}( R_2, 0 )=0$.  Throughout we assume that raw initial non-Clifford states can be prepared with a fidelity that is independent of $\ell$, and these have unit cost, so that $\mathfrak{C}( M_\ell, \epsilon_{\mathrm{raw}} )=1$.  This last assumption is warranted in light of the results of Li~\cite{ying15}.  

\subsection{Gate-synthesis cost}

In general, gate synthesis techniques have an inherent cost that we denote as $\mathfrak{T}(U,\epsilon_{\mathrm{GS}})$, where $\epsilon_{\mathrm{GS}}$ is the precision of the syntheis.   This $T$-count assumes perfect $\ket{M_3}$ magic states are available. Also accounting for the cost of distilled $\ket{M_3}$ states, the full cost is
\begin{equation}
    \mathfrak{C}^{GS}( R_\ell , \eta_\ell ) = \mathfrak{T}( R_\ell , \epsilon_{\mathrm{GS}}) \cdot \mathfrak{C}( M_3 , \epsilon_3)
\end{equation}
where
\begin{equation}
    \eta_\ell \simeq \epsilon_{\mathrm{GS}} + \mathfrak{T}( R_\ell , \epsilon_{\mathrm{GS}}) \cdot \epsilon_3.
\end{equation}
Notice that gate synthesis has an inherent error $\epsilon_{\mathrm{GS}}$ and will also fail if one of the $\mathfrak{T}$ noisy states fail. 
In our analysis, we present data points for combinations of actual instances of gate synthesis using the SR protocol.   The SR data is obtained using the Gridsynth package~\footnote{the Gridsynth package is available at http://www.mathstat.dal.ca/~selinger/newsynth/}.  While SR is optimal for unitary synthesis, lower overheads can be using PQF, which uses ancilla, measurements and feed forward.  For PQF, we use the approximation
\begin{align}
\label{PQF}
	\mathfrak{T}^{PQF}(U, \epsilon_{\mathrm{GS}}) \simeq & \log_2(\sqrt{2}\epsilon_{\mathrm{GS}}^{-1}) \\ \nonumber
 &+4\log_2(\log_2(\sqrt{2}\epsilon_{\mathrm{GS}}^{-1})+1.187,
\end{align}
which is the lowest currently known overhead for gate synthesis. We show in App.~\ref{APP_noise} that the notion of $\epsilon$ used in the gate synthesis literature is comparable to our definition.

\section{Results for small $\ell$}
\label{modestL}

We postpone the technical details on noise analysis until App.~\ref{APP_noise_analysis}, and here present results demonstrating the benefits of MEK$_\ell$.   

\begin{figure*}
\includegraphics{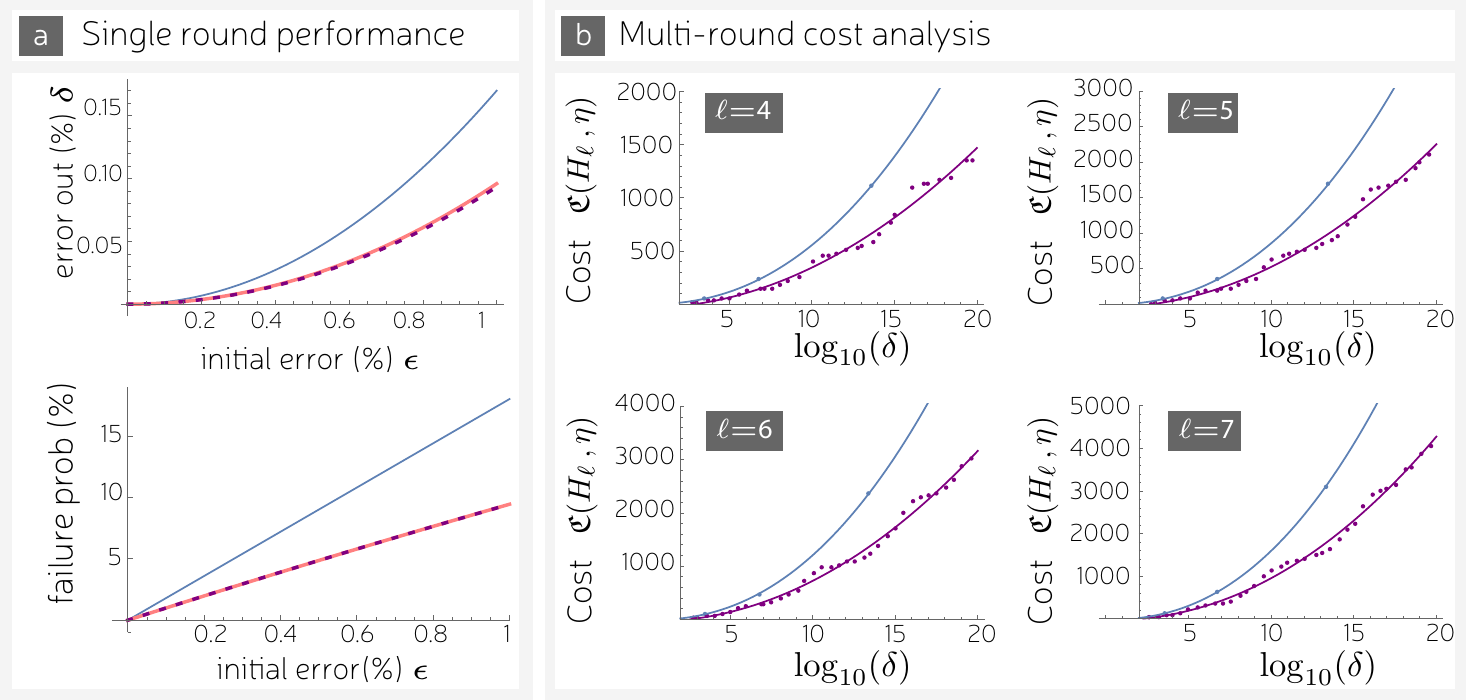}
\caption{Comparison of the resource cost of distillation using MEK$_\ell$ and DP$_\ell$.  (a)  shows performance when $\epsilon_3=\epsilon$, $\epsilon_\ell=\epsilon$ and $\delta=\epsilon^2$.  These are benchmarked against the standard MEK$_3$ protocol (pink), where the curves for MEK$_\ell$ (purple and dotted) are barely distinguishable from MEK$_3$, whereas DP$_\ell$ (blue) performs much worse.  The curves for DP$_\ell$ are based on leading order approximations~\cite{duclosPC}. (b) shows the full resource cost of MEK$_\ell$ protocol (purple) and DP$_\ell$ (blue) of distilling a $\ket{M_\ell}$ state of final error of $\delta$ using resources with an initial error of $1 \%$. Data for DP$_\ell$ taken from Table 1 of Ref.~\cite{duclos15}.  Lines are fitted functions of the form $\mathfrak{C}=a \log(\delta)^b +c$.  }
\label{compareGS}
\end{figure*}

\subsection{Comparison with DP protocol}

\begin{figure*}
\includegraphics{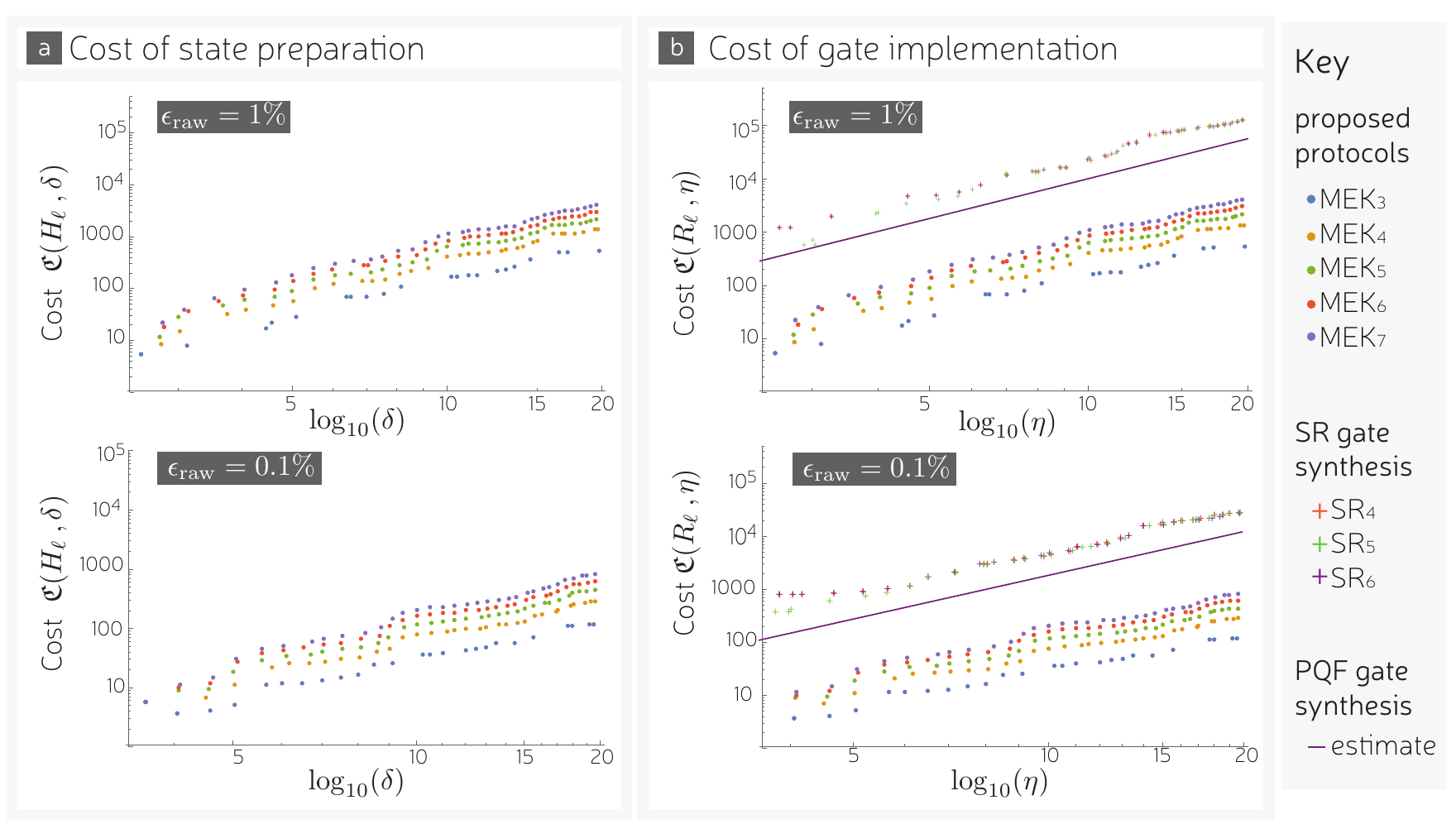}
\caption{The resource cost of MEK$_\ell$ .  (a) shows cost $\mathfrak{C}(\ket{M_\ell},\delta)$ of preparing a magic state $\ket{M_\ell}$ at error rate $\delta$.  (b) shows cost $\mathfrak{C}( R_\ell,\eta)$ of performing a non-Clifford $R_\ell$ at error rate $\eta$. For comparison, (b) also shows the cost of gate-synthesis using the protocol of Ref.~\cite{RS14}. Both (a) and (b) use initial error rates of $\epsilon=0.01$ and $\epsilon=0.001$ for axial rotations of angle $\theta_{\ell}=\pi / 2^{\ell}$ with $\ell=3,4,\ldots 7$.  Smaller rotations and other gate-synthesis protocols are considered later.}  
\label{compareMEK_DP}
\end{figure*}

There are three aspects to the comparison: the number of resources required per attempt, the success probability and how the protocols suppress errors.  Regarding resources, each round of MEK$_\ell$ uses 8 fewer $\ket{M_3}$ states than a round of DP$_\ell$.   In Figs.~(\ref{compareMEK_DP}a) and~(\ref{compareMEK_DP}b) we fix $\epsilon_\ell=\epsilon_3=\epsilon$ and $\eta_{\ell-1}=\epsilon^2$ and compare the performance of MEK$_\ell$ and DP$_\ell$, both benchmarked against MEK$_3$.  We remark that $\eta_{\ell-1}$ must be set significantly lower than other errors as the protocol can not detect noise in the pivotal rotation.  In this context, MEK$_\ell$ is barely indistinguishable from MEK$_3$.  This is expected as the protocols perform identically when $\eta_{\ell-1}=0$, and since $\eta_{\ell-1}=\epsilon^2$ is very small we only observe a very slight difference between MEK$_\ell$ and MEK$_3$.  In contrast, DP$_\ell$ performs worse despite consuming more resources.  

We also consider the full cost of performing many rounds of MEK$_\ell$ and DP$_\ell$.  Because the cost of the pivotal rotation increases with $\ell$, we now see a variation in performance with $\ell$.  Our results are shown for $\ell=4,5,6,7$ in Fig.~(\ref{compareMEK_DP}c) and compare favourably against results reported by Duclos-Cianci and Poulin.   Roughly, we observe a factor $1/2$ reduction in cost, providing a clear cut case for using our compressed  MEK$_\ell$ protocol rather than the original DP$_\ell$ proposal.  Given that all component metrics (resources per round, failure probability and error out) are very favourable towards MEK$_\ell$, one may have expected a more dramatic reduction in cost over DP$_{\ell}$.  One explanation is that both MEK$_\ell$ and DP$_\ell$ require 1 very high fidelity pivotal rotation, which is very costly, and this shared cost limits the extent to which MEK$_\ell$ can outperform DP$_\ell$.

Duclos-Cianci and Poulin discuss a potential improvement to their scheme that uses larger code sizes.  These larger code blocks would encode, and hence distill, $2m$ copies of $\ket{M_\ell}$ and consume $(8m+8)$ copies of $\ket{M_3}$ and $m$ pivotal rotations.  In the large $m$ regime, the ratio of input to output resources (the rate) becomes comparable to MEK${_\ell}$. However, larger code blocks involve more complex circuits and typically do not suppress noise as effectively. Furthermore, moving to large block codes will substantially increase the failure probability, which scales at least linearly with the block size. The exact performance of this large block protocol is unclear and DP presented a rough estimate based on leading order approximations.  For large block codes the number of possible undetected errors of high weight can grow at a combinatorial rate and so it is unclear how accurate leading order approximations will be.    In contrast, MEK$_{\ell}$ offers the improved rate without any of these drawbacks and MEK$_{\ell}$ will also outperform the large block variant of DP$_{\ell}$.  For a precise comparison, one must move beyond leading order approximations and presently no such analysis is available for the large block code variant of DP$_\ell$. 

\subsection{Comparison with gate synthesis protocol for modest $\ell$}

Here, we present our own analysis of gate synthesis for modest $\ell=3,\ldots,7$, with higher levels discussed in following sections.   Fig.~\ref{compareGS} shows a comparison of MEK$_\ell$ against SR and PQF.  For both SR and PQF, the resource cost has little dependence on $\ell$.  In contrast, the resource cost of MEK$_\ell$ increases gradually with $\ell$.  Fig.~(\ref{compareGS}a) reports the magic state cost $\mathfrak{C}(M_\ell, \epsilon_\ell)$. However, for a fair comparison to gate synthesis we must benchmark against the cost of implementing a $R_\ell$ gate, which is $\mathfrak{C}(R_\ell, \eta_\ell)$ in our notation and shown in Fig.~(\ref{compareGS}b).  

We consider the cases of raw error rates of $\epsilon_{\mathrm{raw}}=0.01$ and $\epsilon_{\mathrm{raw}}=0.001$.  The case of higher raw noise is an important benchmark as it has been widely studied~\cite{Bravyi12,Meier13,duclos15}, and here we see improvements over gate synthesis.   For instance, at $\delta = 10^{-15}$ we find PQF is $\sim 22.5$ times more costly than MEK$_{6}$.  However, the lower raw noise regime $\epsilon_{\mathrm{raw}}=0.001$ is in many ways more interesting.  In this regime, the advantage of using MEK$_\ell$ is further improved by a slight margin.  For instance, at target error rate $\delta = 10^{-15}$ we find PQF is $\sim 24$ times more costly than MEK$_{6}$, which is a larger factor than in the high noise regime.  This widening gap between gate synthesis and MEK$_{\ell}$ is seen for all $\ell$ and $\delta$. This increase in gap is intuitive because the distillation cost drops with $\epsilon_{\mathrm{raw}}$, but the $T$-count of gate synthesis is independent of $\epsilon_{\mathrm{raw}}$.
 
While the high noise regime is widely studied, we next argue that the lower noise regime is also more realistic.  Underneath magic state factories is a layer of quantum error correction.  The highest known thresholds for error correction are $ \sim 1 \%$ for the toric code~\cite{wang03,Rauss07,RHG01a,fowler12b} and $ \sim 3 \%$ for Knill's model of postselected quantum computation~\cite{Knill05}.  To prevent astronomical overheads, physical gates must be comfortably below the threshold, and so we assume all physical gates have infidelities well below $1 \%$. Because preparing raw magic states will involve several physical gates, it has often been assumed that $\epsilon_{\mathrm{raw}}$ will be higher than the physical gate error rate, maybe even an order of magnitude higher.  However, Li~\cite{ying15} has shown that we can probabilistically prepare raw magic states at infidelities of about half the physical control-$X$ gate error, assuming control-$X$  failure is the dominant noise mechanism.   We remind the reader that we assume logical level control-$X$ gates are ideal, but beneath the hood of error correction we have very noisy physical control-$X$ gates.  All this indicates that $\epsilon_{\mathrm{raw}}=0.001$ is a feasible regime, more plausible than $\epsilon_{\mathrm{raw}}=0.01$. As such, following subsections focus on the low-noise regime.

\section{Magic state dilution}

\begin{figure}
\includegraphics{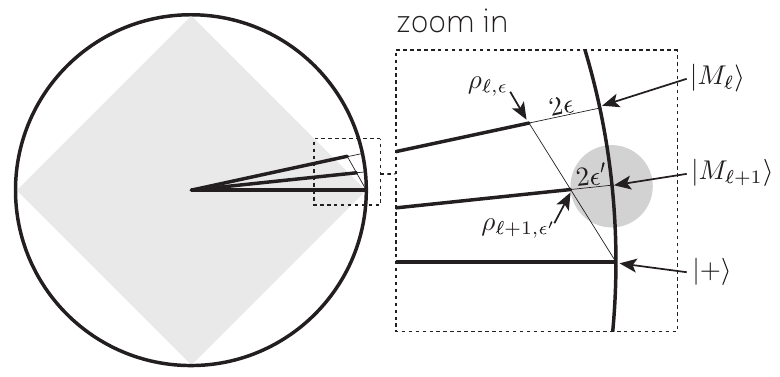}	
\caption{A geometric representation of the dilution protocol in a cross-section of the Bloch sphere.  Note that two points separated by a geometric distance of $d$ in the Bloch sphere, correspond to operators with $d/2$ distance in 1-norm. The grey diamond shows the set of stabiliser states. Given a $\rho_{\ell, \epsilon}$ state and a pure $\ket{+}$ state we can prepare mixtures on the line between these points.  This line intersects the line for states of the form $\rho_{\ell, \epsilon'}$.  We see that for sufficiently large $\epsilon$, the resulting $\epsilon'$ can be reduced $\epsilon' \leq \epsilon$. Furthermore, we show a $2\epsilon'$ radius ball about the point $\ket{M_{\ell+1}}$ to highlight that both $\rho_{\ell, \epsilon}$ state and $\ket{+}$ are further than $2\epsilon'$ away, and so these states would provide a worse approximation of $\ket{M_{\ell+1}}$ than their mixture.}
\label{Fig_dilute}
\end{figure}

As we ascend the Clifford hierarchy, we expect the cost of our protocol to increase.  However, for sufficiently small angles the associated magic states become very close to stabiliser states, and we should be able to exploit this to reduce resource costs. In the analysis of Duclos-Cianci and Poulin, they replaced noisy $\ket{M_\ell}$ states with a stabiliser state whenever $\ell$ exceeded 8. This enabled resource costs to eventually drop with $\ell$.   The $\ket{+}$ state can be considered a noisy $\ket{M_\ell}$ state, though with coherent (non-diagonal) noise. We calculate  the noise of this stabiliser state with respect to $\ket{M_\ell}$ and find
\begin{align}
	 \frac{1}{2}|| (\kb{+}{+} -\kb{M_\ell}{M_\ell}) ||_{1} 
& =|\sin(\theta_{\ell})|  \\ \nonumber
& \sim \theta_{\ell} = \frac{\pi}{2^{\ell}} ,
\end{align}
where the last line gives the small angle approximation. Therefore,  the resource becomes free whenever  $\pi / (2^{\ell})$ is smaller than the target error rate. 

We present an alternative solution that keeps us within the framework of diagonal noise and ensures rapid decrease of costs whenever $\theta_{\ell}^2/\sqrt{2}$ is smaller than the required error rate.  This $\ell$ cutoff is quadratically smaller than that of Duclos-Cianci and Poulin. Let $\rho$ and $\rho'$ be two resource states with costs $\mathfrak{C}$ and $\mathfrak{C}'$.  If we choose to generate $\rho$ with probability $\lambda$ and $\rho'$ with probability $1-\lambda$, then we have the random mixture $\lambda \rho + (1-\lambda)\rho'$.  Furthermore, the expected cost is $\lambda \mathfrak{C} + (1-\lambda) \mathfrak{C}'$.  We consider mixing a noisy $\ket{M_\ell}$ state with a $\ket{+}$ to obtain a good approximation of a noisy $\ket{M_{\ell+1}}$ state.  We say the state $\ket{M_\ell}$ has been diluted, since this allows a source of $\ket{M_\ell}$ states to provide, on average, a greater number of noisy $\ket{M_{\ell+1}}$ states.  We will see that while dilution may increase noise, there are practically relevant regimes where dilution reduces noise.

We use $\rho_{\ell,  \epsilon}$ for a $\ket{M_\ell}$ state with $\epsilon$ diagonal noise, so that
\begin{equation}
	\rho_{\ell,  \epsilon} = (1-\epsilon) \kb{M_{\ell}}{M_{\ell}} + \epsilon \kb{\bar{M}_{\ell}}{\bar{M}_{\ell}}	.
\end{equation}
Preparing $\rho_{\ell,  \epsilon}$ costs $\mathfrak{C}(M_\ell, \epsilon)$ resources. The principle result of this section the relation
\begin{equation}
\label{Eq_dilute}
	\rho_{\ell+1,  \epsilon' } = \lambda \rho_{\ell, \epsilon} + (1-\lambda) \kb{+}{+}	,
\end{equation}
where
\begin{align}
	\lambda   & = \frac{1}{2(1-\epsilon)}	,\\ \label{eq_epsilon_diluted}
	\epsilon' & = \frac{1}{2}\left(1 - (1-2\epsilon) \left[ \frac{\cos(\theta_\ell)}{(1-\epsilon)  } \right] \right).
\end{align}
The dilution provides $\rho_{\ell+1,  \epsilon' }$ at a cost $\lambda \mathfrak{C}(M_\ell, \epsilon) $, which is half the cost of the $\rho_{\ell, \epsilon'}$ state when $\epsilon$ is small.   Remarkably, dilution can even decrease the error rate, provided $\ell$ is large enough. Specifically, if $\ell$ is large enough that the fraction in square bracket exceeds 1, then we have $\epsilon' \leq \epsilon $.  A simpler sufficient condition is $\theta_{\ell} \leq  \sqrt{2\epsilon}$, which can be used to predict a transition in the performance of dilution at 
\begin{equation}
 \ell_c = \log_2 \left( \frac{\pi}{\sqrt{2 \epsilon}} \right).
\end{equation}
In this regime the cost is halved, and iterating this process decreases costs exponentially with $\ell$.  However, even when $\ell$ is does not satisfy the above condition, an increase in error rate may still make dilution more efficient than distillation.  We give a proof of Eq.~(\ref{Eq_dilute}) in App.~\ref{App_dilute} with the underlying geometric intuition presented in Fig.~(\ref{Fig_dilute}).

\section{Results for higher $\ell$}
\label{sec_highL}

\begin{figure*}
\includegraphics{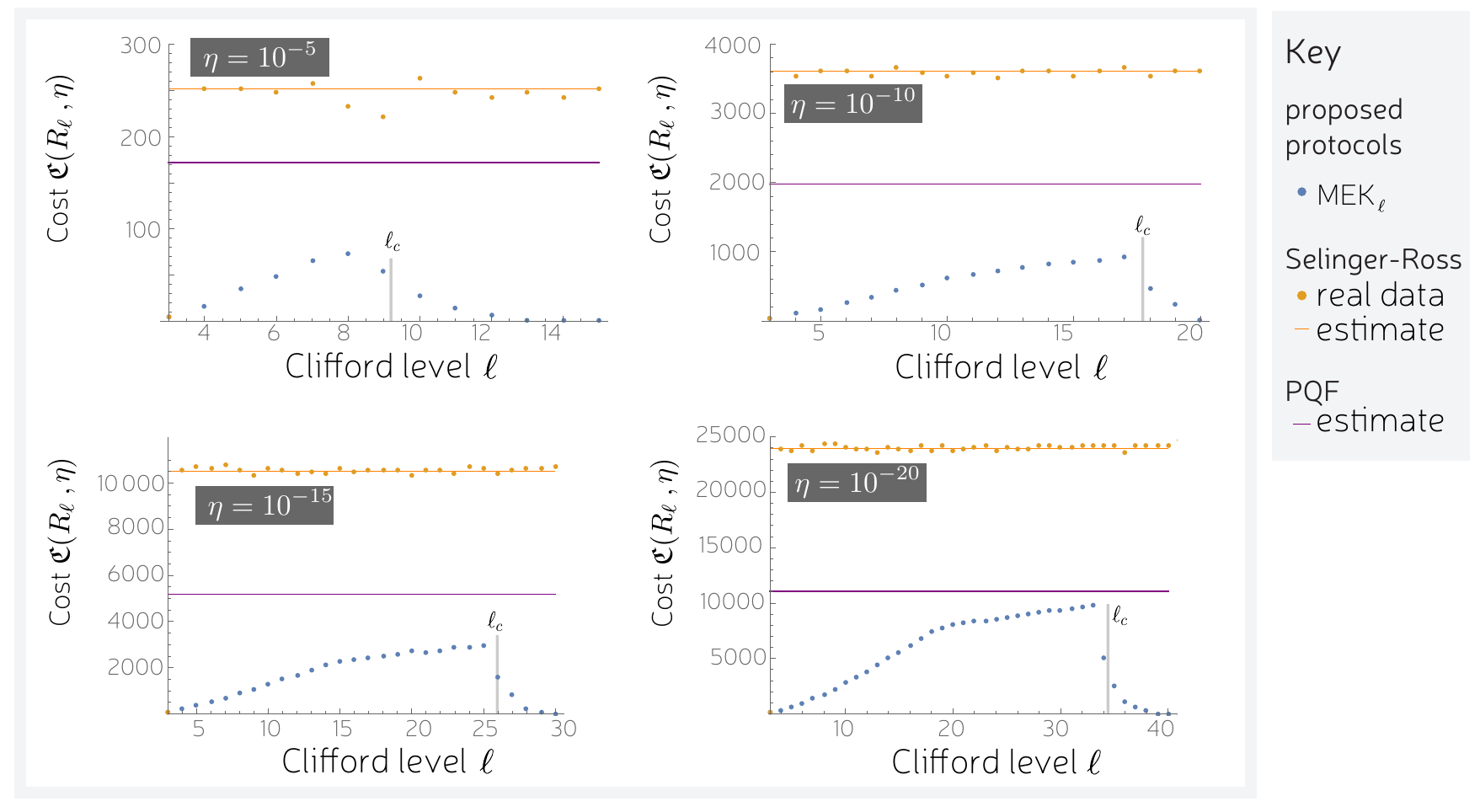}
\caption{The resource cost $R_{\ell}$ rotations using MEK$_\ell$ distillation combined with magic state dilution with $\epsilon_{\mathrm{raw}}=0.1\%$.  Each plot is for a different final error rate $\eta$, and is against the Clifford level $\ell$ beginning with the first nonClifford level $\ell=3$.  We compare perform against gate-synthesis methods, which perform mostly independently of $\ell$.  For the Selinger-Ross (SR) gate synthesis protocol~\cite{RS14}, we show exact data extracted from the gridsynth application and also the analytically proven typical performance.  We also show the typical performance for probabilistic quantum circuits with fallback (PQF)~\cite{bocharov15}.  All results here assume $\epsilon_{\mathrm{raw}}=0.001$.}
\label{highL_fig}
\end{figure*}

Here we discuss the performance of MEK$_{\ell}$ combined with dilution at performing small angle rotations, beyond $\ell=7$. Our results are generated iteratively.  After having compiled a list of achievable costs $\mathfrak{C}(M_\ell, \epsilon)$ and $\mathfrak{C}(R_\ell, \eta)$ for different values of $\epsilon$, we next built lists for $\ell+1$.  The first step is to build a list of $\mathfrak{C}(M_{\ell+1}, \epsilon')$ derived from $\mathfrak{C}(M_{\ell}, \epsilon)$ using the dilution protocol in the previous section.  Next, we considered different combinations of input states into the MEK$_{\ell}$, allowing for using diluted states without any further distillation or inputting diluted states into MEK$_{\ell}$.  The results are presented in Fig.~(\ref{highL_fig}) as a function of $\ell$, showing how resource costs scale with decreasing angle.

For the target error rates $10^{-10},10^{-15}, 10^{-20}$ we can see three clear regions, with the middle region absent for the $10^{-5}$ plot.  First, the resource cost increases roughly linearly with $\ell$. Next, there is a transition where the gradient becomes gentler. Lastly, at some cutoff the cost starts to fall exponentially with $\ell$.  In this last regime, we rely solely on diluting magic states.  This exponential cliff was predicted by the analysis in the previous section, and is labeled in the plots by $\ell_c$.  The behaviour of the middle region is also due to dilution. Here we typically find that one round of MEK$_\ell$ is used, with the input noisy $\ket{M_{\ell}}$ states produced by dilution, as opposed to using two or more rounds of MEK$_\ell$ on a raw resource of error rate $\epsilon_{\mathrm{raw}}$.  Since dilution is less effective at low $\ell$, in this regime we are completely reliant on MEK$_{\ell}$ and here observe the most rapid increase in costs.

Our results are also presented against the cost of two gate-synthesis methods SR and PQF.  We see that for small $\ell$, and large $\ell > \ell_c$ there is a significant gain over both gate-synthesis methods by over an order of magnitude.  In the intermediate regime, our protocol still outperforms gate-synthesis but approaches a similar order of magnitude.

\section{Conclusions}

We have proposed a protocol for distilling magic states providing small angle rotations at a lower cost than the previous proposal of Duclos-Cianci and Poulin.  The cost, as measured by number of raw magic states used, is less than best current gate synthesis techniques.  For modest size rotation of angle $\pi/(2^6)$, we saw our protocol operates $\sim 24$ times better than gate synthesis.  Assessing this improvement, a factor 12 can be attributed to the innovations of Duclos-Cianci and Poulin, with our compression of the protocol providing the additional factor 2.  To perform smaller angle rotations, the resource cost gradually increases until the resources become close to stabiliser states.  To maintain an advantage over gate-synthesis this phenomenon must be exploited.  We proposed a magic state dilution protocol, which  converts one magic state into a larger number suitable for a smaller angle rotation.  This exponentially suppresses resource costs for angles $\theta_{\ell} \leq  \sqrt{2\epsilon}$.  In this regime, resource costs are many orders of magnitude lower than gate-synthesis.  

Throughout, we have quantified cost by raw magic states consumed.  This is the standard metric in every paper that has introduced a new protocol for magic state distillation.  Other quantities of  interest are the total number of qubits used (space cost) and depth of circuits used  (time costs).  These full resource costs include stabiliser and Clifford costs, but are highly sensitive to the specific architecture used.  The fully costed performance of Bravyi-Haah and Reed-Muller magic state distillation have been considered for surface code architectures, both in a braiding picture~\cite{fowler13} and using transversal logical gates~\cite{Gorman12}.  Such full resource assessments are significant undertakings that followed the initial proposals, and so lie beyond our present scope.  However, all such analyses to date have found that protocols with lower $T$-counts also have lower full resource costs. Although, in Ref.~\cite{fowler13} they noted that significant improvements in protocol $T$ counts can lead to more modest improvements of full resource costs.  This occurs because the surface code incurs a polylogarthmic overhead in $\epsilon^{-1}$ that grows more rapidly than overheads due to magic state distillation or gate-synthesis.  That is, in a fully costed analysis the surface code overhead dwarfs all other overheads.  Though this peculiarity could disappear if future developments provided a more efficient and practical alternative over the surface code.

Here we have considered only single-qubit gate-synthesis, but in parallel work there has been progress on unifying magic state distillation with gate-synthesis for a class of multi-qubit circuits~\cite{campbell16,campbell16b}. We close by remarking that the fields of magic state distillation and gate synthesis have rapidly evolved in the last several years, giving good reason to be hopeful that advancement in these areas will continue to bring quantum computation ever closer to reality.  

\begin{figure*}
    \includegraphics{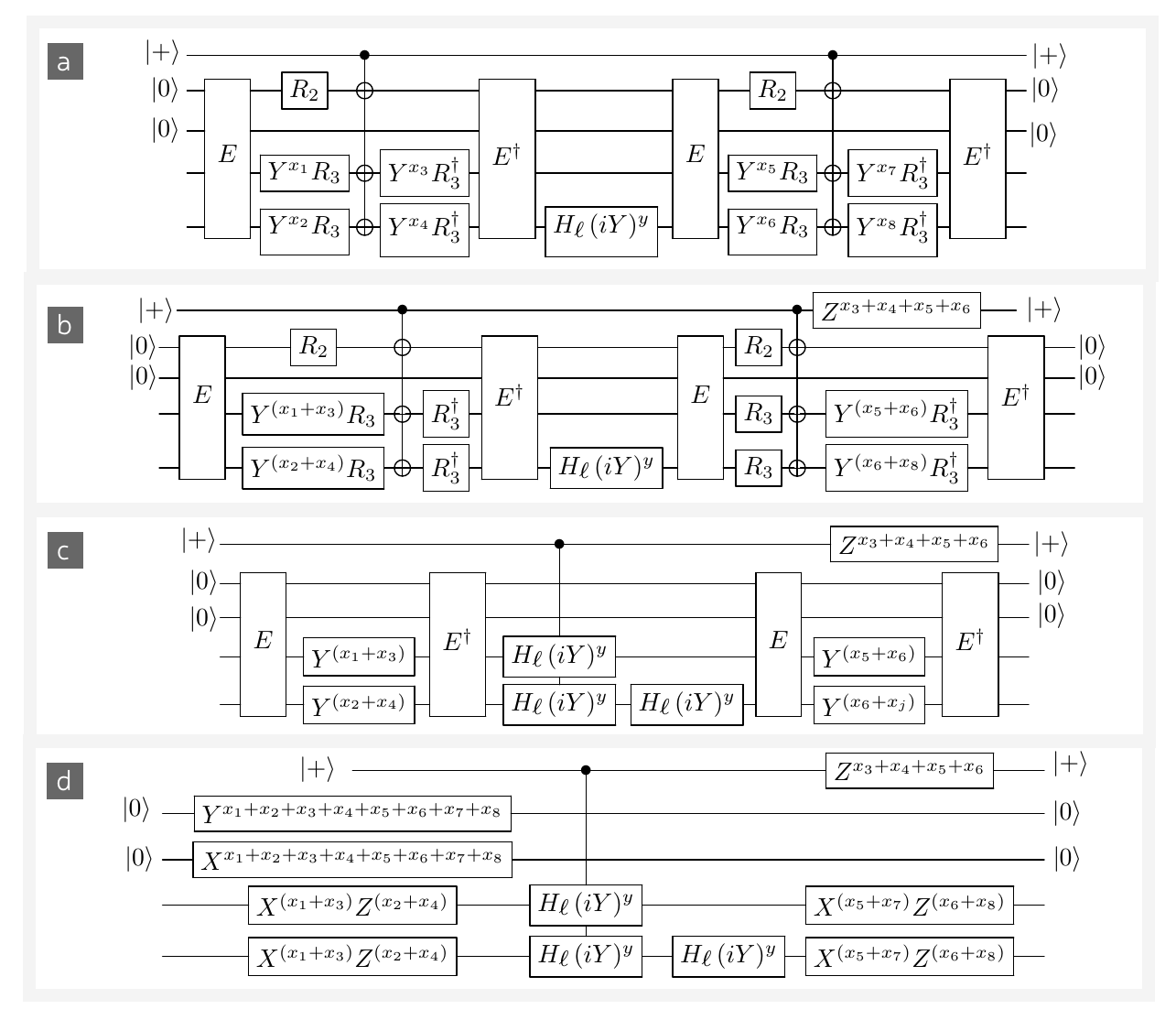}
    \caption{Propagation noise terms around distillation circuit.}
    \label{FIG_noise}
\end{figure*}

\section{Acknowledgements}

This work was supported by the EPSRC (grant EP/M024261/1) and collaboration between the authors was facilitated by the NQIT quantum hub.  We thank Bryan Eastin and Adam Meier for assistance with reproducing the simulation of MEK. We thank Guillaume Duclos-Cianci and Mark Howard for comments on the manuscript.  Gate synthesis calculations used the Gridsynth package, and we thank Peter Selinger and Neil J. Ross for writing this software and making it available.  We thank the developers of the Qcircuit package used in generation of figures. All our analysis is available in the Supplementary material as the Mathematica source files and also in pdf format.

\appendix

\section{Arbitrary angle rotations}
\label{Arb_angle}

We focus on small angle rotations, but some of our techniques also apply to larger rotations.  In particular, not only is $R_\ell$ in the $\ell^{\mathrm{th}}$ level of the Clifford hierarchy, but $R_\ell^m$ belongs to the same level for any integer $m$.  For the compressed MEK$_{\ell}$ protocols described we can replace $R_\ell \rightarrow R_\ell^m$ throughout and they perform identically, and so $R_\ell^m$ has the same resource cost as $R_\ell$.  

 If we require some rotation with an angle $\phi$ that is not a rational fraction of $\pi$, then it will not fall within any level of the Clifford  hierarchy.  Rather, we may resort to inexact synthesis and find a $n\theta_\ell$ that is sufficiently close to $\phi$.  The larger we choose $\ell$, the better the approximation, but the higher the resource cost.  Specifically, for all $\phi$ and $\ell$ we can always find an $n$ such that $| \phi - n\theta_{\ell} | \leq \pi / 2^{\ell}$, which improves exponentially with $\ell$.  To prove this, we observe that for fixed $\ell$ the set of reachable angles $\{ n \theta_\ell  |  1 \leq n \leq 2^{\ell +1} \}$ are equally spaced on the interval $[ 0, 2\pi ]$.  Therefore, the reachable angles have gaps between them of distance $\pi/2^{\ell}$. The angle $\phi$ furthest from any angle in the reachable set will sit halfway between two reachable angles, and so this worst case $\phi$ lies $\pi/2^{\ell+1}$ away from the nearest reachable angle. For instance, to ensure $| \phi - n\theta_{\ell} | \leq 10^{-10}$ we can use $\ell=35$.  It is straightforward to verify that high precision in angle ensures closeness (upto a constant factor) in diamond norm distance between the corresponding channels.  

However, for large $\ell$ using MEK$_{\ell}$ becomes increasingly costly with $\ell$.
For large $\ell$, this is overcome by using magic state dilution.  This protocol relies on $\ket{M_\ell}$ being close to the $\ket{+}$ stabiliser state to reduce overheads, and so the above comments about replacing $R_\ell \rightarrow R_\ell^m$ do not apply to dilution.  Therefore,  the resource costs presented in Sec.~\ref{sec_highL} only apply when $\phi$ is close to some $\theta_{\ell}$.

\section{Precision in gate-synthesis}
\label{APP_noise}

Across the gate-synthesis literature, different notions of precision are used.  Here we discuss how these relate to the diamond norm measure of noise. We assume throughout that all unitaries are in $SU(2)$ and so have determinant 1.

In Def 7.1 of Ref.~\cite{RS14}, SR state that they quantify precision using the spectral norm $\epsilon_{SR}:=||U-V ||_{\infty}$.   Let $W=U^\dagger V$ and have eigenvalues $\exp(i \varphi)$ and $\exp(-i \varphi)$ with $\varphi \geq 0$, so that $\varphi$ is small whenever $U$ and $V$ are close. It follows that $|| U - V ||_{\infty} =|1-\exp(i \varphi)|= 2 |\sin(\frac{\varphi}{2})|\sim \varphi $.

If $\mathcal{U}$ and $\mathcal{V}$ are channels associated with $U$ and $V$, then Refs.~\cite{Wang13,Wang15}  show the inequality
\begin{align}
\epsilon_{GS}:=	\frac{1}{2}||\mathcal{U} - \mathcal{V} ||_{\diamond} & \leq || U - V ||_{\infty} \sim \varphi.
\end{align}
We now consider a lower bound.  For any $\rho$ with $||\rho||_1=1$ we have by definition that 
\begin{align}
	\frac{1}{2}||\mathcal{U} - \mathcal{V} ||_{\diamond} & \geq \frac{1}{2}|| (\mathcal{U}\otimes \id)\rho - (\mathcal{V}\otimes \id)\rho ||_{1} .
\end{align}
Setting $\rho= \kb{+}{+} \otimes \kb{+}{+}$, we find that 
\begin{align}
	\frac{1}{2}||\mathcal{U} - \mathcal{V} ||_{\diamond} & \geq \frac{1}{2} |1-\exp(i 2 \varphi)| \\ \nonumber
 & = | \sin(  \varphi) | \sim \varphi
\end{align}
Therefore, in the small $\varphi$ limit we have $\epsilon_{GS}=\epsilon_{SR}$.

In the work on PQF, they measure precision~\footnote{Private communication with Alex Bocharov} as follows
\begin{equation}
		\epsilon_{PQF} := \sqrt{1 - \frac{1}{2}\mathrm{tr}(U^\dagger V) }.
\end{equation}
This evaluates to $\epsilon_{PQF} = \sqrt{2} |\sin (\varphi / 2)| \sim \varphi / \sqrt{2}$.  Therefore, this quantity is $\sqrt{2}$ smaller than our diamond norm measure, so that $\epsilon_{PQF}=\epsilon_{GS}/ \sqrt{2}$.  This adjustment is shown in Eq.~\eqref{APP_noise}, which carries a $\sqrt{2}$ correction relative to the result of Ref.~\cite{bocharov15}.  Although we have carefully accounted for differences in precision measures, this small constant factor adjustment in error rate makes a negligible difference to resource overheads.

\section{Noise analysis}
\label{APP_noise_analysis}

This section shows how the circuit operates when the non-Clifford gates ($R_3$ and $R_{\ell-1}$) are noisy due to imperfect magic states.   When a $R_\ell$ gate fails due to noise, it is followed by an additional $Y$ rotation.  There are 8 locations that errors can occur on $R_3$ gates and we define a binary vector $x=(x_1,x_2,x_3,x_4,x_5,x_6,x_7,x_8)$ that records whether a $Y$ error occurs at a particular $R_3$ gate. We use the binary variable $y=0,1$ to track an error in the pivotal rotation $R_{\ell-1}$.  When the $R_{\ell-1}$ gate fails, we have a rotation $H_\ell(iY)$ rather than $H_\ell$.  Notice the complex phase $i$, which makes no physical difference to the unitary. However, aspects of the proof rely on $H_\ell$ being Hermitian, and the additional phase keeps it Hermitian. The resulting random circuit is shown in Fig.~(\ref{FIG_noise}a).  

We pull some $Y$ noise operators through the control phase gates, which can cause $Z$ noise on the control as shown in Fig.~(\ref{FIG_noise}b).  The central portion of the circuit now contains no noise terms (except $y$) and so we can replace it with a logical control ($H_\ell \otimes H_\ell$) or its equivalent when $y=1$. This yields Fig.~(\ref{FIG_noise}c).

Next, we pull the $Y$ noise backwards through the encoder $E$.  We already know how $E$ acts on Pauli operators from which we conclude that the inverse action satisfies
\begin{align}
    E^\dagger : &  Y_3 \rightarrow Y_1 X_2 Z_3 Z_4 ,\\
    E^\dagger : &  Y_4 \rightarrow Y_1 X_2 X_3 X_4 .
\end{align}
Similarly random unitaries $Y_3^\alpha Y_4^\beta $ map as 
\begin{align}
    E^\dagger : &  Y_3^\alpha Y_4^\beta \rightarrow Y_1^{\alpha+\beta} X_2^{\alpha+\beta} ( Z_3^\alpha X_3^\beta )( Z_4^\alpha X_4^\beta )
\end{align}
Applying this rule to our circuit yields Fig.~(\ref{FIG_noise}d).

From Fig.~(\ref{FIG_noise}d), we see that to obtain the correct measurement outcome on qubit 1 or 2 requires that 
\begin{equation}
\label{condition}
|x|:= x_1+x_2+x_3+x_4+x_5+x_6+x_7+x_8 = 0 \pmod 2,  
\end{equation} which detects any single error on the $R_3$ gates. This shows that the error correction properties of our original circuit have survived the compression.  

From this circuit we can get some intuition for the leading order error terms presented in Eqs.~\ref{delta_out}.  First consider the failure probability, $P_{\mathrm{fail}}=1-P_{\mathrm{suc}}$.  If $|x|:= x_1+x_2+x_3+x_4+x_5+x_6+x_7+x_8 = 1 \pmod 2$, then we detect an error and have a failure.  The leading order contribution is single error processes, and there are 8 such errors, so this contributes $8 \epsilon_3$ to $P_{\mathrm{fail}}$. Otherwise, if $x=(0,0,\ldots 0)$ and the pivotal rotation also works, then we perform a perfect projection onto the $+ H_{\ell} \otimes H_\ell$ subspace.  This projection detects an errors, and the protocol declares failure if one of the two noisy $\ket{M_{\ell}}$ states is faulty, which occurs with probability $2\epsilon_\ell$.  Lastly,  we consider if everything works except the pivotal rotation.  The pivotal rotation fails with probability $\eta_{\ell -1}$, but not all this contributes to the failure probability.  When the pivotal rotation fails, and the rest of the circuit works correctly, it results in the operation shown in Fig.~(\ref{FIG_IDENTs1}c), but with the replacement $H_\ell \rightarrow i H_\ell Y$.  Algebraically, this operation is
\begin{align}
    V &= \frac{1}{2}\left[ \id + (i H_\ell Y )\otimes (i H_\ell Y ) \right] [(i H_\ell Y)  \otimes \id )] \\ \nonumber
     &= \frac{-i}{2}\left[ (Y H_\ell )  \otimes \id +  \id \otimes (Y H_\ell   ) \right] 
\end{align}
Applying this channel to $\ket{M_\ell}\ket{M_\ell}$, and noting that $Y \ket{M_\ell} = \ket{\bar{M}_\ell}$, we have
\begin{equation}
\label{piv_fail}
  \ket{\psi}=  V \ket{M_\ell}\ket{M_\ell} =\frac{-i}{2}( \ket{\bar{M}_\ell}\ket{M_\ell} + \ket{M_\ell} \ket{\bar{M}_\ell} )
\end{equation}
Therefore, the probability of an even parity measurement, combined with pivotal rotation failure, is $ \eta_{\ell -1} \bk{\psi}{\psi}$.  Recalling, $\bk{M_\ell}{\bar{M}_\ell}=0$, we find $\bk{\psi}{\psi}=1/2$.
Therefore, the failure probability gains a contribution of $\frac{1}{2}\eta_{\ell -1} $.  This covers all the leading order processes that contribute to a detected failure.

Next we consider leading order processes that go undetected, but output an erroneous state $\ket{\bar{M}_\ell}$.  The protocol outputs a two qubit state, and we trace out the second qubit and evaluate the error rate on the first qubit.  Switching the first and second output qubit will yield the same result. First we consider pairs of errors in the $R_3$ gates.  There are 28 such pairs, all satisfying $|x|=0\pmod{2}$. However, the parity measurement must yield the correct outcome also.  Since we are only considering leading order errors, we can assume the noisy $\ket{M_\ell}$ states are error free.  Therefore, to obtain the $\ket{+}$ on the control ancilla, there cannot be a $Z$ flip on the control, and so $x_3+x_4+x_5+x_6=0 \pmod 2$.  This cuts the number of undetected error pairs down to 14. Not all undetected error pairs lead to a logical error on the first qubit.  For $x_1=x_2=1$ and $x_7=x_8=1$ we see that there is a logical $Y \otimes Y$ error, and so both these processes contribute.  For the other 12 undetected errors, the parity projection becomes deformed so that it projects onto a state that on average has $1/2$ overlap with $\ket{M_\ell}$.  Let us expand on this notion of a deformed parity projection by considering the case $x_1 = x_5 =1$, with other combinations proving similar. This causes an $X \otimes X$ rotation both before and after the parity projection, so that instead of projecting onto $H_\ell \otimes H_\ell$ we project onto $H^\perp_\ell \otimes H^\perp_\ell$ where $H^\perp_\ell := X H_\ell X$, which we call a deformed projection.  How much this particular process contributes varies with $\theta_\ell$, but a lengthily evaluation over all such error pairs shows the average contribution is $1/2$. This totals the error contribution from failed $R_3$ gates to $(2 + \frac{12}{2}) \epsilon_3^2 = 8 \epsilon_3^2$.  If the $R_3$ gates do not fail, but instead we have a perfect parity projection, then a logical error occurs if the projection is applied to $\ket{\bar{M}_\ell}\ket{\bar{M}_\ell}$, which occurs with probability $\epsilon_\ell^2$. Lastly, we contemplate when the pivotal rotation cairres an error. We see from Eqn.~(\ref{piv_fail}) that $|\bk{\bar{M}_\ell , M_\ell}{\psi}|=1/4$ and so the this leads to a logical error with probability $\frac{1}{4}\delta_{\ell-1}$.  We remark again that the leading order contribute here is linear, and not quadratically suppressed, and so the pivotal rotation must be high fidelity to enable distillation.

\section{Generic noise}

Throughout the main text, we assumed diagonal noise as described in Sec.~\ref{quant_noise}.  Here we show that MEK$_{\ell}$ is robust against generic noise, making use of the 1-norm and diamond norm to measure error rates.  Specifically, our proof will demonstrate that MEK$_{\ell}$ noise is quadratically reduced, converging towards zero.  Our proof will provide an upper bound on the output error rate, rather than the exact expression. 

We first consider MEK$_{\ell}$ as a quantum channel $\mathcal{E}$ acting on two noisy $\rho_{\ell, \epsilon_\ell}$ states, such that
\begin{equation}
	\epsilon_\ell := 	\frac{1}{2}|| \rho_{\ell, \epsilon_\ell} -  \rho_{\ell} ||_1.
\end{equation}
where here $\rho_{\ell, \epsilon_\ell}$ may not be diagonal in the $\{ \ket{M_\ell },  \ket{\bar{M}_\ell } \}$ basis.   For brevity, we use $\rho_{\ell}:=\rho_{\ell,0}=\kb{M_{\ell}}{M_{\ell}}$.  Below, we also use $\Delta_\ell := \rho_{\ell, \epsilon_\ell} -  \rho_{\ell}$ where $||\Delta_\ell||_1=2\epsilon_\ell$.

The protocol uses 8 $\rho_{3, \epsilon_3}$ states, which by Clifford twirling arguments can be forced to have purely diagonal noise.  We use $x =\{ x_1, \ldots x_8\}$ to label define Pauli errors from the $\rho_{3,\epsilon_3}$ states, so that
\begin{equation}
		\rho_{3,\epsilon_3}^{\otimes 8} = \sum_{x \in \mathbb{Z}_2^8 } p_x Y[x] \rho_{3}^{\otimes 8} Y[x], 
\end{equation}
where $Y[x]:=\otimes_{j} Y_j^{x_j}$ and 
\begin{equation}
	p_x := \epsilon_3^{|x|}(1-\epsilon_3)^{8-|x|},	
\end{equation}
with $|x|:=\sum_j x_j$.  For a given error configuration $x$, the protocol implements $\mathcal{E}_x$.  We decompose this as  $\mathcal{E}_x = \mathcal{E}''_x \circ \mathcal{P}_{\ell-1} \circ \mathcal{E}'_x$, where $\mathcal{P}_{\ell-1}$ is the noisy pivotal rotation, $ \mathcal{E}'_x$ is the circuit before the pivotal rotation and $\mathcal{E}''_x$ is the circuit afterwards.  Throughout, we use the symbol $\circ$ to denote composition of channels. Lastly we trace out one of the qubits, using the partial trace map $\ptr$, to give the single qubit output.  The trace is over either qubit $a=1$ or $a=2$, and our analysis will be independent of this choice.

The complete channel is
\begin{equation}
    \mathcal{E} = \sum_{x} p_x \ptr \circ \mathcal{E}''_x \circ \mathcal{P}_{\ell-1} \circ \mathcal{E}'_x.
\end{equation}
Next, we use that since $\frac{1}{2}||\mathcal{P}_{\ell-1}-\mathcal{U}_{\ell-1}||_{\diamond}=\eta_{\ell-1}$, we know $\mathcal{P}_{\ell-1}=\mathcal{U}_{\ell-1} +  \mathcal{D}$, where $||\mathcal{D}||_\diamond=2 \eta_{\ell-1}$. Note that $\mathcal{D}$ is not necessarily a positive map.  This entails
\begin{align}
    \mathcal{E} = & \left( \sum_{x} p_x \ptr \circ \mathcal{E}''_x \circ \mathcal{U}_{\ell-1} \circ \mathcal{E}'_x \right)   \\ \nonumber
& +   \left( \sum_{x} p_x \ptr \circ \mathcal{E}''_x \circ \mathcal{D} \circ \mathcal{E}'_x \right)  .
\end{align}   
Recall from Eq.~\eqref{condition} that the terms $\mathcal{E}''_x \circ \mathcal{U}_{\ell-1} \circ \mathcal{E}'_x$ vanish whenever $x$ has an odd number of 1s.  For the nonvanishing $x$ values, we split the first bracket into two components, representing the $x=(0,0,\ldots,0)$ term and then the nonzero $|x|=2,4,6,8$ terms
\begin{align}
    \mathcal{E} = &   p_0 \ptr \circ \mathcal{E}''_0 \circ \mathcal{U}_{\ell-1} \circ \mathcal{E}'_0 \\ \nonumber
 & + \left( \sum_{|x|=2,4,6,8} p_x \ptr \circ \mathcal{E}''_x \circ \mathcal{U}_{\ell-1} \circ \mathcal{E}'_x \right)  \\ \nonumber
 & + \eta_{\ell-1}  \left( \sum_{x} p_x \ptr \circ \mathcal{E}''_x \circ \mathcal{D} \circ \mathcal{E}'_x \right)  .
\end{align}   
We next introduce some new notation to simplify this expression to
\begin{align}
    \mathcal{E} & =   \mathcal{A}  +   \mathcal{B} + \mathcal{C} .
\end{align} 
where our new maps are
\begin{align} 
    \mathcal{A} & = \mathrm{tr}_a  \circ p_0 \mathcal{E}''_0 \circ \mathcal{U}_{\ell-1} \circ \mathcal{E}'_0 , \\
    \mathcal{B} & = \mathrm{tr}_a \circ  \left( \sum_{|x|=2,4,6,8} p_x \mathcal{E}''_x \circ \mathcal{U}_{\ell-1} \circ \mathcal{E}'_x \right) , \\
    \mathcal{C} & = \mathrm{tr}_a \circ   \sum_{x} p_x \mathcal{E}''_x \circ \mathcal{D} \circ \mathcal{E}'_x .
\end{align}
One can easily verify the channels satisfy
\begin{align}
    || \mathcal{A} ||_{\diamond} & \leq p_g  :=  p_0 = (1-\epsilon_3)^8 , \\
    || \mathcal{B} ||_{\diamond} & \leq p_b  := \sum_{x=2,4,6,8} \binom{8}{x} \epsilon_3^{|x|} (1 - \epsilon_3)^{8-|x|} , \\
    || \mathcal{C} ||_{\diamond} & \leq 2 \eta_{\ell-1} .
\end{align}
Later we also make use of (assuming $\epsilon_3 \leq 0.01$) the following\begin{align}
\label{pgbInequals}
    1- 8 \epsilon_3   \leq & p_g \leq 1 \\ 
        28 \epsilon_3^2 - 168 \epsilon_3^3  \leq & p_b \leq 28 \epsilon_3^2 \end{align}
We next consider the effect of these channels individually before composing the results together.

The $\mathcal{A}$ map represents a perfect parity projection with no errors, so that
\begin{align}
\label{eqA00}
    \mathcal{A} ( \rho_{\ell} \otimes \rho_{\ell} ) = p_g \ptr [ \rho_{\ell} \otimes \rho_{\ell} ] = p_g \rho_{\ell}.
\end{align}
We consider its action on a noisy state $\rho_{\ell,\epsilon_{\ell}} = \rho_{\ell} + \Delta_\ell$.  In the $\ket{M_\ell}$ basis we have
\begin{equation}
    \Delta_\ell = \epsilon_\ell \left( \begin{array}{cc} - a & b \\
 b^* & a 
 \end{array} \right).
\end{equation}
From  $||\Delta_\ell||=2\epsilon_\ell$, we deduce $0 \leq |a|^2+|b|^2 \leq 1$.  Performing the parity projection on single error terms yields
\begin{align}
\label{eqA01}
     \mathcal{A} ( \rho_\ell \otimes \Delta_\ell ) &=-p_g a\epsilon_{\ell} \ptr [ ( \rho_\ell  \otimes \rho_\ell  ) ] =- p_g \epsilon_{\ell} a \rho_\ell  \\
     \mathcal{A} ( \Delta_\ell \otimes \rho_\ell   ) &=-p_g a \epsilon_\ell \ptr [ -( \rho_\ell  \otimes \rho_\ell  ) ] =-p_g a \epsilon_\ell \rho_\ell  
\end{align}
Lastly for the double error term we have 
\begin{align}
\label{eqA11}
     \mathcal{A} ( \Delta_\ell \otimes \Delta_\ell ) =&p_g a^2 \epsilon_\ell^2[  \rho_\ell +(Y \rho_\ell Y) ]
\end{align}
Combining Eq.~(\ref{eqA00}), Eq.~(\ref{eqA01}) and Eq.~(\ref{eqA11}), we have that 
\begin{equation}
     \mathcal{A} ( \rho_\ell \otimes \rho_\ell  )   =p_g [(1- a \epsilon_\ell)^2\rho_\ell + (a \epsilon_\ell)^2 (Y \rho_\ell Y)] , 
\end{equation}
where $-1 \leq a \leq 1$.

The $\mathcal{B}$ map represents when the pivotal rotation works perfectly, but at least two $\rho_{\ell, \epsilon_{3}}$ states cause an error.  By elementary norm properties we have
\begin{equation}
   || \mathcal{B}( \rho_{\ell, \epsilon_\ell} \otimes \rho_{\ell, \epsilon_\ell}) ||_1 \leq || \mathcal{B} ||_\diamond \cdot || \rho_{\ell, \epsilon_\ell} \otimes \rho_{\ell, \epsilon_\ell} ||_1 \leq p_b.
\end{equation}
We define $\rho_\mathcal{B} := \mathcal{B}( \rho_\ell \otimes \rho_\ell) $, and since  $\mathcal{B}$ is a completely positive channel we have that $\mathrm{tr}[ \rho_\mathcal{B} ] = || \rho_{\mathcal{B}} ||_1$.

Lastly, the $\mathcal{C}$ map represents a failed pivotal rotation. We define $\rho_\mathcal{C} := \mathcal{C}( \rho_{\ell, \epsilon_\ell} \otimes \rho_{\ell, \epsilon_\ell})  $ and use norm properties to deduce $||\rho_\mathcal{C}||_1 \leq 2 \eta_{\ell-1}$.  Since $\mathcal{C}$ is not a positive map, all we know regarding the trace is that  $- 2 \eta_{\ell-1} \leq \mathrm{tr}[\rho_\mathcal{C}] \leq 2 \eta_{\ell-1}$.

Putting all these components together we have that
\begin{align}
    \mathcal{E}[ \rho_{\ell, \epsilon_\ell} \otimes \rho_{\ell, \epsilon_\ell}] =&  p_g[(1- a \epsilon_\ell)^2 \rho_\ell + (a \epsilon_\ell)^2 (Y \rho_\ell Y)]  \\ \nonumber
 &+ \rho_\mathcal{B} + \rho_\mathcal{C}
\end{align}
Taking the trace gives the success probability
\begin{equation}
    P_{\mathrm{suc}} =  p_g(1- 2a \epsilon_\ell+2a^2\epsilon_{\ell}^2) +  \mathrm{tr}[\rho_\mathcal{B}]  +  \mathrm{tr}[\rho_\mathcal{C}] .
\end{equation}
The smallest value this can take is when $\mathrm{tr}[\rho_\mathcal{B}]=0$, $\mathrm{tr}[\rho_\mathcal{C}]=-2 \eta_{\ell-1}$, and $a=1$.  This gives the rigorous, albeit pessimistic bound, that  
\begin{equation}
    P_{\mathrm{suc}} \geq  p_g(1- 2 \epsilon_\ell+ 2 \epsilon_{\ell}^2) - 2 \eta_{\ell-1} .
\end{equation}
We renormalise to obtain
\begin{align}
  \rho^{\mathrm{out}} & = \frac{1}{P_{\mathrm{suc}}}  \mathcal{E}[\rho_{\ell, \epsilon_\ell} \otimes \rho_{\ell, \epsilon_\ell}].
\end{align}
The output error rate $\delta_\ell :=\frac{1}{2} ||\rho^{\mathrm{out}} - \rho_\ell||_1$ is then
\begin{align}
    \delta_\ell =  \frac{1}{ 2P_{\mathrm{suc} } }& \bigg| \bigg| [p_g(1- a \epsilon_\ell)^2 -P_{\mathrm{suc}}]\rho_\ell   \\ \nonumber
 & + p_g(a \epsilon_\ell)^2 (Y \rho_\ell Y) + \rho_\mathcal{B} + \rho_\mathcal{C} \bigg| \bigg|_1.
\end{align}
Using the triangle inequality we obtain
\begin{align}
    \delta_\ell \leq &  \frac{ |p_g(1- a \epsilon_\ell)^2 -P_{\mathrm{suc}}| + p_g|a \epsilon_\ell |^2 + ||\rho_\mathcal{B}|| + ||\rho_\mathcal{C}||  }{2P_{\mathrm{suc}}} \\ \nonumber
&  \leq \frac{  p_g a^2 \epsilon_\ell^2 + ||\rho_\mathcal{B}|| + ||\rho_\mathcal{C}||  }{P_{\mathrm{suc}}} \\ \nonumber
&  \leq \frac{  p_g a^2 \epsilon_\ell^2 + p_b + 2 \eta_{\ell-1}  }{P_{\mathrm{suc}}}
\end{align}
The worst case scenario is again that $a=1$, so that 
\begin{equation}
     \delta_\ell \leq \frac{ p_g \epsilon_\ell^2 + p_b + 2 \eta_{\ell-1}}{p_g(1- 2 \epsilon_\ell+ 2  \epsilon_\ell^2) - 2 \eta_{\ell-1}}.
\end{equation}
Using the worst case bound on $p_g$ and $p_b$ from Eq.~\eqref{pgbInequals}, we have 
\begin{equation}
     \delta_\ell \leq  \frac{ \epsilon_\ell^2 + 28 \epsilon_3^2 + 2 \eta_{\ell-1}}{(1-8\epsilon_3^2)(1- 2 \epsilon_\ell+ 2\epsilon_\ell^2) - 2 \eta_{\ell-1}} \sim \epsilon_\ell^2 + 28 \epsilon_3^2 + 2 \eta_{\ell-1} ,
\end{equation}
where the last approximation gives the leading order terms, showing quadratic suppression in $\epsilon_\ell$ and $\epsilon_3$.  This completes our proof that MEK$_{\ell}$ suppresses generic noise.

We remark that though quadratic the prefactor of $\epsilon_3$ is 3.5 times larger than in Eq.~\eqref{delta_out} and Eq.~\eqref{delta_out_exact}, and the prefactor of $\eta_{\ell-1}$ is 8 times larger.  However, these increased prefactors are mainly artefacts of approximations made in the proof.  We report that we also found a proof with a leading order approximation $\epsilon_\ell^2 + 8 \epsilon_3^2 +  \eta_{\ell-1}$, though the proof is much longer.

\section{Dilution proof}
\label{App_dilute}

Here we verify the dilution expression in Eq.~\eqref{Eq_dilute}.  As an intermediately step in our proof, we begin by considering the mixture $\frac{1}{2}\left( \rho_{\ell}+ \kb{+}{+} \right)$.  The pure magic state $\rho_{\ell}$ can be expanded in the Pauli basis as
\begin{align}
	\rho_{\ell}  & = \frac{1}{2} ( \id + H_\ell) \\ \nonumber
& = \frac{1}{2}\left( \id + \cos(\theta_{\ell-1})X+\sin(\theta_{\ell-1})Z \right),	
\end{align}
and $\kb{+}{+}=\frac{1}{2}(\id + X)$.  Therefore, in the Pauli basis
\begin{align}
\frac{\left( \rho_{\ell}+ \kb{+}{+} \right)}{2}	 = \frac{1}{2} \left( \id +  q(\cos(\phi)X+ \sin(\phi)Z  ) \right),
\end{align}
where we introduce the variables $q$ and $\phi$ as follows
\begin{align}
	q & = \frac{1}{2}\sqrt{(\cos(\theta_{\ell-1})+1)^2 + \sin(\theta_{\ell-1})^2}, \\ \nonumber
	\sin(\phi) & = \frac{\sin(\theta_{\ell-1})}{2 q}.
\end{align}
Using standard trigonometric identities, we find  
\begin{align}
	q & = \cos( \theta_\ell ), \\ \nonumber
	\phi & = 2 \theta_{\ell-1} = \theta_{\ell}.
\end{align}
This entails
\begin{align}
\label{eq_purecase}
\frac{ \rho_{\ell}+ \kb{+}{+} }{2}	& = \cos(\theta_\ell) \rho_{\ell+1} + (1-\cos(\theta_\ell))\frac{\id}{2} , \\ \nonumber
& = \rho_{\ell+1, \sin^2 (\theta_{\ell+1})}.
\end{align}
where in the last line we have used $(1-\cos(\theta_\ell))/2 = \sin^2 (\theta_{\ell+1})$. Let us recap.  By 50-50 mixing a pure magic state $\ket{M_{\ell}}$ with a pure $\ket{+}$ state, we obtain a noisy $\ket{M_{\ell+1}}$ state with error rate $\sin^2 (\theta_{\ell+1})$.  Next, we extend the proof to account for noise on the initial magic state. 

Given a noisy magic state $\rho_{\ell, \epsilon}$, this can be decomposed as
\begin{equation}
	\rho_{\ell, \epsilon} = (1-2\epsilon)	\rho_{\ell} +  \epsilon \id.
\end{equation}
Mixing this with the $\ket{+}$ stabiliser state by an amount $\lambda$, gives
\begin{equation}
	\lambda \rho_{\ell, \epsilon} + (1-\lambda)\kb{+}{+}= \lambda(1-2\epsilon)\rho_{\ell} + (1-\lambda)\kb{+}{+}+ \lambda \epsilon \id.
\end{equation}
We choose $\lambda$ so that the coefficients of $\rho_{\ell}$ and $\kb{+}{+}$ are equal, which requires
\begin{equation}
	\lambda = \frac{1}{2(1-\epsilon)}.	
\end{equation}
This yields
\begin{align}
	& \lambda \rho_{\ell, \epsilon} + (1-\lambda)\kb{+}{+} \\ \nonumber
	& = \frac{(1-2\epsilon)}{1-\epsilon}\left[ \frac{ \rho_{\ell} + \kb{+}{+} }{2} \right]+ \frac{\epsilon \id}{2(1-\epsilon)}.
\end{align}
We can now apply Eq.~\eqref{eq_purecase} to the terms in the square brackets, so that
\begin{align}
	& \lambda \rho_{\ell, \epsilon} + (1-\lambda)\kb{+}{+}  \\ \nonumber
&= \frac{\cos(\theta_\ell)(1-2\epsilon)}{1-\epsilon}\rho_{\ell+1, 0}+ \frac{(1-\cos(\theta_\ell))(1-2\epsilon)+\epsilon }{2(1-\epsilon)}\id,
\end{align}
which can be more compactly written as
\begin{align}
	 \lambda \rho_{\ell, \epsilon} + (1-\lambda)\kb{+}{+}  &= \rho_{\ell+1, \frac{1-q'}{2}},
\end{align}
where
\begin{equation}
	q' = \cos(\theta_\ell) \frac{1-2\epsilon}{1-\epsilon} = \cos(\theta_\ell) \frac{1-2\epsilon}{1-\epsilon}.	
\end{equation}
The most relevant quantity to report is the error rate $\epsilon' = (1-q')/2$, which follows immediately from the above as given by Eq.~\eqref{eq_epsilon_diluted}.  

\section{Simulation results}
\label{APP_EXPRESS}

Using Mathematica, we symbolically simulate the circuit in Fig.~(\ref{FIG_noise}d).  Full details are available in the Supplementary material (see MEKL\textunderscore simulation.nb).

We find the output states are again diagonal in the $\{ \ket{M_\ell} ,  \ket{\bar{M}_\ell} \}$ basis.  The main text quoted the leading order error and success probability as Eqs.~(\ref{delta_out}), and use we present the exact results here
\begin{widetext}
\begin{align}
\label{delta_out_exact}
    \delta_\ell   = \frac{1}{P_\mathrm{suc}} \bigg( & 8 \epsilon_3^2 + \epsilon_\ell^2 + \frac{1}{4}\eta_{\ell-1}  
  - 2 \eta_{\ell-1} \epsilon_3 +   6 \eta_{\ell-1} \epsilon_3^2 - 48 \epsilon_3^3 -   8 \eta_{\ell-1} \epsilon_3^3 + 136 \epsilon_3^4 +   4 \eta_{\ell-1} \epsilon_3^4 - 224 \epsilon_3^5 + 224 \epsilon_3^6 \\ \nonumber
  &- 128 \epsilon_3^7 +   32 \epsilon_3^8 + \epsilon_\ell^2 - \eta_{\ell-1} \epsilon_\ell^2 -   8 \epsilon_3 \epsilon_\ell^2 + 
  8 \eta_{\ell-1} \epsilon_3 \epsilon_\ell^2 +   24 \epsilon_3^2 \epsilon_\ell^2 -   24 \eta_{\ell-1} \epsilon_3^2 \epsilon_\ell^2 -   32 \epsilon_3^3 \epsilon_\ell^2 + 32 \eta_{\ell-1} \epsilon_3^3 \epsilon_\ell^2 \\ \nonumber
  &+   16 \epsilon_3^4 \epsilon_\ell^2 - 16 \eta_{\ell-1} \epsilon_3^4 \epsilon_\ell^2  \bigg) , \\
   P_\mathrm{suc}    = &  448 \epsilon_3^5 - 448 \epsilon_3^6 + 256 \epsilon_3^7 -  64 \epsilon_3^8 + 
 1/2 \eta_{\ell-1} (1 - 2 \epsilon_3)^4 (1 - 2 \epsilon_\ell)^2 +  2 \epsilon_\ell - 2 \epsilon_\ell^2 \\ \nonumber
 &+ 64 \epsilon_3^3 (2 - \epsilon_\ell + \epsilon_\ell^2) - 32 \epsilon_3^4 (9 - \epsilon_\ell + \epsilon_\ell^2) +  8 \epsilon_3 (1 - 2 \epsilon_\ell + 2 \epsilon_\ell^2) -  8 \epsilon_3^2 (5 - 6 \epsilon_\ell + 6 \epsilon_\ell^2).
\end{align}
\end{widetext}


\begin{thebibliography}{49}%
\makeatletter
\providecommand \@ifxundefined [1]{%
 \@ifx{#1\undefined}
}%
\providecommand \@ifnum [1]{%
 \ifnum #1\expandafter \@firstoftwo
 \else \expandafter \@secondoftwo
 \fi
}%
\providecommand \@ifx [1]{%
 \ifx #1\expandafter \@firstoftwo
 \else \expandafter \@secondoftwo
 \fi
}%
\providecommand \natexlab [1]{#1}%
\providecommand \enquote  [1]{``#1''}%
\providecommand \bibnamefont  [1]{#1}%
\providecommand \bibfnamefont [1]{#1}%
\providecommand \citenamefont [1]{#1}%
\providecommand \href@noop [0]{\@secondoftwo}%
\providecommand \href [0]{\begingroup \@sanitize@url \@href}%
\providecommand \@href[1]{\@@startlink{#1}\@@href}%
\providecommand \@@href[1]{\endgroup#1\@@endlink}%
\providecommand \@sanitize@url [0]{\catcode `\\12\catcode `\$12\catcode
  `\&12\catcode `\#12\catcode `\^12\catcode `\_12\catcode `\%12\relax}%
\providecommand \@@startlink[1]{}%
\providecommand \@@endlink[0]{}%
\providecommand \url  [0]{\begingroup\@sanitize@url \@url }%
\providecommand \@url [1]{\endgroup\@href {#1}{\urlprefix }}%
\providecommand \urlprefix  [0]{URL }%
\providecommand \Eprint [0]{\href }%
\providecommand \doibase [0]{http://dx.doi.org/}%
\providecommand \selectlanguage [0]{\@gobble}%
\providecommand \bibinfo  [0]{\@secondoftwo}%
\providecommand \bibfield  [0]{\@secondoftwo}%
\providecommand \translation [1]{[#1]}%
\providecommand \BibitemOpen [0]{}%
\providecommand \bibitemStop [0]{}%
\providecommand \bibitemNoStop [0]{.\EOS\space}%
\providecommand \EOS [0]{\spacefactor3000\relax}%
\providecommand \BibitemShut  [1]{\csname bibitem#1\endcsname}%
\let\auto@bib@innerbib\@empty
\bibitem [{\citenamefont {Eastin}\ and\ \citenamefont
  {Knill}(2009)}]{Eastin09}%
  \BibitemOpen
  \bibfield  {author} {\bibinfo {author} {\bibfnamefont {B.}~\bibnamefont
  {Eastin}}\ and\ \bibinfo {author} {\bibfnamefont {E.}~\bibnamefont {Knill}},\
  }\href@noop {} {\bibfield  {journal} {\bibinfo  {journal} {Phys. Rev. Lett.}\
  }\textbf {\bibinfo {volume} {102}},\ \bibinfo {pages} {110502} (\bibinfo
  {year} {2009})}\BibitemShut {NoStop}%
\bibitem [{\citenamefont {Kitaev}\ \emph {et~al.}(2002)\citenamefont {Kitaev},
  \citenamefont {Shen},\ and\ \citenamefont {Vyalyi}}]{kitaev02}%
  \BibitemOpen
  \bibfield  {author} {\bibinfo {author} {\bibfnamefont {A.~Y.}\ \bibnamefont
  {Kitaev}}, \bibinfo {author} {\bibfnamefont {A.}~\bibnamefont {Shen}}, \ and\
  \bibinfo {author} {\bibfnamefont {M.~N.}\ \bibnamefont {Vyalyi}},\
  }\href@noop {} {\emph {\bibinfo {title} {Classical and quantum
  computation}}},\ Vol.~\bibinfo {volume} {47}\ (\bibinfo  {publisher}
  {American Mathematical Society Providence},\ \bibinfo {year}
  {2002})\BibitemShut {NoStop}%
\bibitem [{\citenamefont {Bravyi}\ and\ \citenamefont
  {Kitaev}(2005)}]{BraKit05}%
  \BibitemOpen
  \bibfield  {author} {\bibinfo {author} {\bibfnamefont {S.}~\bibnamefont
  {Bravyi}}\ and\ \bibinfo {author} {\bibfnamefont {A.}~\bibnamefont
  {Kitaev}},\ }\href@noop {} {\bibfield  {journal} {\bibinfo  {journal} {Phys.
  Rev. A}\ }\textbf {\bibinfo {volume} {71}},\ \bibinfo {pages} {022316}
  (\bibinfo {year} {2005})}\BibitemShut {NoStop}%
\bibitem [{\citenamefont {Meier}\ \emph {et~al.}(2013)\citenamefont {Meier},
  \citenamefont {Eastin},\ and\ \citenamefont {Knill}}]{Meier13}%
  \BibitemOpen
  \bibfield  {author} {\bibinfo {author} {\bibfnamefont {A.~M.}\ \bibnamefont
  {Meier}}, \bibinfo {author} {\bibfnamefont {B.}~\bibnamefont {Eastin}}, \
  and\ \bibinfo {author} {\bibfnamefont {E.}~\bibnamefont {Knill}},\
  }\href@noop {} {\bibfield  {journal} {\bibinfo  {journal} {Quant. Inf. and
  Comp.}\ }\textbf {\bibinfo {volume} {13}},\ \bibinfo {pages} {195} (\bibinfo
  {year} {2013})}\BibitemShut {NoStop}%
\bibitem [{\citenamefont {Bravyi}\ and\ \citenamefont {Haah}(2012)}]{Bravyi12}%
  \BibitemOpen
  \bibfield  {author} {\bibinfo {author} {\bibfnamefont {S.}~\bibnamefont
  {Bravyi}}\ and\ \bibinfo {author} {\bibfnamefont {J.}~\bibnamefont {Haah}},\
  }\href {\doibase 10.1103/PhysRevA.86.052329} {\bibfield  {journal} {\bibinfo
  {journal} {Phys. Rev. A}\ }\textbf {\bibinfo {volume} {86}},\ \bibinfo
  {pages} {052329} (\bibinfo {year} {2012})}\BibitemShut {NoStop}%
\bibitem [{\citenamefont {Jones}(2013{\natexlab{a}})}]{Jones13}%
  \BibitemOpen
  \bibfield  {author} {\bibinfo {author} {\bibfnamefont {C.}~\bibnamefont
  {Jones}},\ }\href {\doibase 10.1103/PhysRevA.87.042305} {\bibfield  {journal}
  {\bibinfo  {journal} {Phys. Rev. A}\ }\textbf {\bibinfo {volume} {87}},\
  \bibinfo {pages} {042305} (\bibinfo {year} {2013}{\natexlab{a}})}\BibitemShut
  {NoStop}%
\bibitem [{\citenamefont {Fowler}\ \emph {et~al.}(2013)\citenamefont {Fowler},
  \citenamefont {Devitt},\ and\ \citenamefont {Jones}}]{fowler13}%
  \BibitemOpen
  \bibfield  {author} {\bibinfo {author} {\bibfnamefont {A.~G.}\ \bibnamefont
  {Fowler}}, \bibinfo {author} {\bibfnamefont {S.~J.}\ \bibnamefont {Devitt}},
  \ and\ \bibinfo {author} {\bibfnamefont {C.}~\bibnamefont {Jones}},\
  }\href@noop {} {\bibfield  {journal} {\bibinfo  {journal} {Scientific
  reports}\ }\textbf {\bibinfo {volume} {3}},\ \bibinfo {pages} {1939}
  (\bibinfo {year} {2013})}\BibitemShut {NoStop}%
\bibitem [{\citenamefont {Kliuchnikov}\ \emph {et~al.}(2013)\citenamefont
  {Kliuchnikov}, \citenamefont {Maslov},\ and\ \citenamefont
  {Mosca}}]{kliuchnikov13}%
  \BibitemOpen
  \bibfield  {author} {\bibinfo {author} {\bibfnamefont {V.}~\bibnamefont
  {Kliuchnikov}}, \bibinfo {author} {\bibfnamefont {D.}~\bibnamefont {Maslov}},
  \ and\ \bibinfo {author} {\bibfnamefont {M.}~\bibnamefont {Mosca}},\
  }\href@noop {} {\bibfield  {journal} {\bibinfo  {journal} {Physical review
  letters}\ }\textbf {\bibinfo {volume} {110}},\ \bibinfo {pages} {190502}
  (\bibinfo {year} {2013})}\BibitemShut {NoStop}%
\bibitem [{\citenamefont {Paetznick}\ and\ \citenamefont
  {Svore}(2014)}]{paetznick14}%
  \BibitemOpen
  \bibfield  {author} {\bibinfo {author} {\bibfnamefont {A.}~\bibnamefont
  {Paetznick}}\ and\ \bibinfo {author} {\bibfnamefont {K.~M.}\ \bibnamefont
  {Svore}},\ }\href@noop {} {\bibfield  {journal} {\bibinfo  {journal} {Quantum
  Information \& Computation}\ }\textbf {\bibinfo {volume} {14}},\ \bibinfo
  {pages} {1277} (\bibinfo {year} {2014})}\BibitemShut {NoStop}%
\bibitem [{\citenamefont {Gosset}\ \emph {et~al.}(2014)\citenamefont {Gosset},
  \citenamefont {Kliuchnikov}, \citenamefont {Mosca},\ and\ \citenamefont
  {Russo}}]{gosset14}%
  \BibitemOpen
  \bibfield  {author} {\bibinfo {author} {\bibfnamefont {D.}~\bibnamefont
  {Gosset}}, \bibinfo {author} {\bibfnamefont {V.}~\bibnamefont {Kliuchnikov}},
  \bibinfo {author} {\bibfnamefont {M.}~\bibnamefont {Mosca}}, \ and\ \bibinfo
  {author} {\bibfnamefont {V.}~\bibnamefont {Russo}},\ }\href@noop {}
  {\bibfield  {journal} {\bibinfo  {journal} {Quantum Information \&
  Computation}\ }\textbf {\bibinfo {volume} {14}},\ \bibinfo {pages} {1261}
  (\bibinfo {year} {2014})}\BibitemShut {NoStop}%
\bibitem [{\citenamefont {Ross}\ and\ \citenamefont {Selinger}(2014)}]{RS14}%
  \BibitemOpen
  \bibfield  {author} {\bibinfo {author} {\bibfnamefont {N.~J.}\ \bibnamefont
  {Ross}}\ and\ \bibinfo {author} {\bibfnamefont {P.}~\bibnamefont
  {Selinger}},\ }\href@noop {} {\bibfield  {journal} {\bibinfo  {journal}
  {arXiv preprint arXiv:1403.2975}\ } (\bibinfo {year} {2014})}\BibitemShut
  {NoStop}%
\bibitem [{\citenamefont {Amy}\ and\ \citenamefont {Mosca}(2016)}]{amy16}%
  \BibitemOpen
  \bibfield  {author} {\bibinfo {author} {\bibfnamefont {M.}~\bibnamefont
  {Amy}}\ and\ \bibinfo {author} {\bibfnamefont {M.}~\bibnamefont {Mosca}},\
  }\href@noop {} {\bibfield  {journal} {\bibinfo  {journal} {arXiv preprint
  arXiv:1601.07363}\ } (\bibinfo {year} {2016})}\BibitemShut {NoStop}%
\bibitem [{\citenamefont {Anderson}\ \emph {et~al.}(2014)\citenamefont
  {Anderson}, \citenamefont {Duclos-Cianci},\ and\ \citenamefont
  {Poulin}}]{Anderson14}%
  \BibitemOpen
  \bibfield  {author} {\bibinfo {author} {\bibfnamefont {J.~T.}\ \bibnamefont
  {Anderson}}, \bibinfo {author} {\bibfnamefont {G.}~\bibnamefont
  {Duclos-Cianci}}, \ and\ \bibinfo {author} {\bibfnamefont {D.}~\bibnamefont
  {Poulin}},\ }\href {\doibase 10.1103/PhysRevLett.113.080501} {\bibfield
  {journal} {\bibinfo  {journal} {Phys. Rev. Lett.}\ }\textbf {\bibinfo
  {volume} {113}},\ \bibinfo {pages} {080501} (\bibinfo {year}
  {2014})}\BibitemShut {NoStop}%
\bibitem [{\citenamefont {Bombin}(2013)}]{bombin13b}%
  \BibitemOpen
  \bibfield  {author} {\bibinfo {author} {\bibfnamefont {H.}~\bibnamefont
  {Bombin}},\ }\href@noop {} {\bibfield  {journal} {\bibinfo  {journal} {arXiv
  preprint arXiv:1311.0879}\ } (\bibinfo {year} {2013})}\BibitemShut {NoStop}%
\bibitem [{\citenamefont {Bombin}\ and\ \citenamefont
  {Martin-Delgado}(2006)}]{bombin06}%
  \BibitemOpen
  \bibfield  {author} {\bibinfo {author} {\bibfnamefont {H.}~\bibnamefont
  {Bombin}}\ and\ \bibinfo {author} {\bibfnamefont {M.~A.}\ \bibnamefont
  {Martin-Delgado}},\ }\href@noop {} {\bibfield  {journal} {\bibinfo  {journal}
  {Physical review letters}\ }\textbf {\bibinfo {volume} {97}},\ \bibinfo
  {pages} {180501} (\bibinfo {year} {2006})}\BibitemShut {NoStop}%
\bibitem [{\citenamefont {Bombin}\ \emph {et~al.}(2013)\citenamefont {Bombin},
  \citenamefont {Chhajlany}, \citenamefont {Horodecki},\ and\ \citenamefont
  {Martin-Delgado}}]{Bombin13}%
  \BibitemOpen
  \bibfield  {author} {\bibinfo {author} {\bibfnamefont {H.}~\bibnamefont
  {Bombin}}, \bibinfo {author} {\bibfnamefont {R.~W.}\ \bibnamefont
  {Chhajlany}}, \bibinfo {author} {\bibfnamefont {M.}~\bibnamefont
  {Horodecki}}, \ and\ \bibinfo {author} {\bibfnamefont {M.~A.}\ \bibnamefont
  {Martin-Delgado}},\ }\href {http://stacks.iop.org/1367-2630/15/i=5/a=055023}
  {\bibfield  {journal} {\bibinfo  {journal} {New Journal of Physics}\ }\textbf
  {\bibinfo {volume} {15}},\ \bibinfo {pages} {055023} (\bibinfo {year}
  {2013})}\BibitemShut {NoStop}%
\bibitem [{\citenamefont {Raussendorf}\ \emph {et~al.}(2007)\citenamefont
  {Raussendorf}, \citenamefont {Harrington},\ and\ \citenamefont
  {Goyal}}]{RHG01a}%
  \BibitemOpen
  \bibfield  {author} {\bibinfo {author} {\bibfnamefont {R.}~\bibnamefont
  {Raussendorf}}, \bibinfo {author} {\bibfnamefont {J.}~\bibnamefont
  {Harrington}}, \ and\ \bibinfo {author} {\bibfnamefont {K.}~\bibnamefont
  {Goyal}},\ }\href@noop {} {\bibfield  {journal} {\bibinfo  {journal} {New
  Journal of Physics}\ }\textbf {\bibinfo {volume} {9}},\ \bibinfo {pages}
  {199} (\bibinfo {year} {2007})},\ \Eprint
  {http://arxiv.org/abs/quant-ph/0703143} {quant-ph/0703143} \BibitemShut
  {NoStop}%
\bibitem [{\citenamefont {O'Gorman}\ and\ \citenamefont
  {Campbell}(2016)}]{Gorman12}%
  \BibitemOpen
  \bibfield  {author} {\bibinfo {author} {\bibfnamefont {J.}~\bibnamefont
  {O'Gorman}}\ and\ \bibinfo {author} {\bibfnamefont {E.~T.}\ \bibnamefont
  {Campbell}},\ }\href@noop {} {\bibfield  {journal} {\bibinfo  {journal}
  {arXiv preprint arXiv:1605.07197}\ } (\bibinfo {year} {2016})}\BibitemShut
  {NoStop}%
\bibitem [{\citenamefont {Brown}\ \emph {et~al.}(2016)\citenamefont {Brown},
  \citenamefont {Nickerson},\ and\ \citenamefont {Browne}}]{brown15}%
  \BibitemOpen
  \bibfield  {author} {\bibinfo {author} {\bibfnamefont {B.~J.}\ \bibnamefont
  {Brown}}, \bibinfo {author} {\bibfnamefont {N.~H.}\ \bibnamefont
  {Nickerson}}, \ and\ \bibinfo {author} {\bibfnamefont {D.~E.}\ \bibnamefont
  {Browne}},\ }\href {http://dx.doi.org/10.1038/ncomms12302} {\bibfield
  {journal} {\bibinfo  {journal} {Nat Commun}\ }\textbf {\bibinfo {volume} {7}}
  (\bibinfo {year} {2016})}\BibitemShut {NoStop}%
\bibitem [{\citenamefont {Wang}\ \emph {et~al.}(2003)\citenamefont {Wang},
  \citenamefont {Harrington},\ and\ \citenamefont {Preskill}}]{wang03}%
  \BibitemOpen
  \bibfield  {author} {\bibinfo {author} {\bibfnamefont {C.}~\bibnamefont
  {Wang}}, \bibinfo {author} {\bibfnamefont {J.}~\bibnamefont {Harrington}}, \
  and\ \bibinfo {author} {\bibfnamefont {J.}~\bibnamefont {Preskill}},\
  }\href@noop {} {\bibfield  {journal} {\bibinfo  {journal} {Annals of
  Physics}\ }\textbf {\bibinfo {volume} {303}},\ \bibinfo {pages} {31}
  (\bibinfo {year} {2003})}\BibitemShut {NoStop}%
\bibitem [{\citenamefont {Raussendorf}\ and\ \citenamefont
  {Harrington}(2007)}]{Rauss07}%
  \BibitemOpen
  \bibfield  {author} {\bibinfo {author} {\bibfnamefont {R.}~\bibnamefont
  {Raussendorf}}\ and\ \bibinfo {author} {\bibfnamefont {J.}~\bibnamefont
  {Harrington}},\ }\href {\doibase 10.1103/PhysRevLett.98.190504} {\bibfield
  {journal} {\bibinfo  {journal} {Phys. Rev. Lett.}\ }\textbf {\bibinfo
  {volume} {98}},\ \bibinfo {pages} {190504} (\bibinfo {year}
  {2007})}\BibitemShut {NoStop}%
\bibitem [{\citenamefont {Fowler}\ \emph {et~al.}(2012)\citenamefont {Fowler},
  \citenamefont {Whiteside}, \citenamefont {McInnes},\ and\ \citenamefont
  {Rabbani}}]{fowler12b}%
  \BibitemOpen
  \bibfield  {author} {\bibinfo {author} {\bibfnamefont {A.~G.}\ \bibnamefont
  {Fowler}}, \bibinfo {author} {\bibfnamefont {A.~C.}\ \bibnamefont
  {Whiteside}}, \bibinfo {author} {\bibfnamefont {A.~L.}\ \bibnamefont
  {McInnes}}, \ and\ \bibinfo {author} {\bibfnamefont {A.}~\bibnamefont
  {Rabbani}},\ }\href@noop {} {\bibfield  {journal} {\bibinfo  {journal}
  {Physical Review X}\ }\textbf {\bibinfo {volume} {2}},\ \bibinfo {pages}
  {041003} (\bibinfo {year} {2012})}\BibitemShut {NoStop}%
\bibitem [{\citenamefont {Anwar}\ \emph {et~al.}(2012)\citenamefont {Anwar},
  \citenamefont {Campbell},\ and\ \citenamefont {Browne}}]{Anwar12}%
  \BibitemOpen
  \bibfield  {author} {\bibinfo {author} {\bibfnamefont {H.}~\bibnamefont
  {Anwar}}, \bibinfo {author} {\bibfnamefont {E.~T.}\ \bibnamefont {Campbell}},
  \ and\ \bibinfo {author} {\bibfnamefont {D.~E.}\ \bibnamefont {Browne}},\
  }\href {http://stacks.iop.org/1367-2630/14/i=6/a=063006} {\bibfield
  {journal} {\bibinfo  {journal} {New Journal of Physics}\ }\textbf {\bibinfo
  {volume} {14}},\ \bibinfo {pages} {063006} (\bibinfo {year}
  {2012})}\BibitemShut {NoStop}%
\bibitem [{\citenamefont {Campbell}\ \emph {et~al.}(2012)\citenamefont
  {Campbell}, \citenamefont {Anwar},\ and\ \citenamefont
  {Browne}}]{Campbell12}%
  \BibitemOpen
  \bibfield  {author} {\bibinfo {author} {\bibfnamefont {E.~T.}\ \bibnamefont
  {Campbell}}, \bibinfo {author} {\bibfnamefont {H.}~\bibnamefont {Anwar}}, \
  and\ \bibinfo {author} {\bibfnamefont {D.~E.}\ \bibnamefont {Browne}},\
  }\href {\doibase 10.1103/PhysRevX.2.041021} {\bibfield  {journal} {\bibinfo
  {journal} {Phys. Rev. X}\ }\textbf {\bibinfo {volume} {2}},\ \bibinfo {pages}
  {041021} (\bibinfo {year} {2012})}\BibitemShut {NoStop}%
\bibitem [{\citenamefont {Campbell}(2014)}]{campbell14}%
  \BibitemOpen
  \bibfield  {author} {\bibinfo {author} {\bibfnamefont {E.~T.}\ \bibnamefont
  {Campbell}},\ }\href@noop {} {\bibfield  {journal} {\bibinfo  {journal}
  {Phys. Rev. Lett.}\ }\textbf {\bibinfo {volume} {113}},\ \bibinfo {pages}
  {230501} (\bibinfo {year} {2014})}\BibitemShut {NoStop}%
\bibitem [{\citenamefont {Dawkins}\ and\ \citenamefont
  {Howard}(2015)}]{dawkins15}%
  \BibitemOpen
  \bibfield  {author} {\bibinfo {author} {\bibfnamefont {H.}~\bibnamefont
  {Dawkins}}\ and\ \bibinfo {author} {\bibfnamefont {M.}~\bibnamefont
  {Howard}},\ }\href@noop {} {\bibfield  {journal} {\bibinfo  {journal}
  {Physical review letters}\ }\textbf {\bibinfo {volume} {115}},\ \bibinfo
  {pages} {030501} (\bibinfo {year} {2015})}\BibitemShut {NoStop}%
\bibitem [{\citenamefont {Landahl}\ and\ \citenamefont
  {Cesare}(2013)}]{landahl13}%
  \BibitemOpen
  \bibfield  {author} {\bibinfo {author} {\bibfnamefont {A.~J.}\ \bibnamefont
  {Landahl}}\ and\ \bibinfo {author} {\bibfnamefont {C.}~\bibnamefont
  {Cesare}},\ }\href@noop {} {\bibfield  {journal} {\bibinfo  {journal} {arXiv
  preprint arXiv:1302.3240}\ } (\bibinfo {year} {2013})}\BibitemShut {NoStop}%
\bibitem [{\citenamefont {Gottesman}\ and\ \citenamefont
  {Chuang}(1999{\natexlab{a}})}]{gottesman99}%
  \BibitemOpen
  \bibfield  {author} {\bibinfo {author} {\bibfnamefont {D.}~\bibnamefont
  {Gottesman}}\ and\ \bibinfo {author} {\bibfnamefont {I.~L.}\ \bibnamefont
  {Chuang}},\ }\href@noop {} {\bibfield  {journal} {\bibinfo  {journal}
  {Nature}\ }\textbf {\bibinfo {volume} {402}},\ \bibinfo {pages} {390}
  (\bibinfo {year} {1999}{\natexlab{a}})}\BibitemShut {NoStop}%
\bibitem [{\citenamefont {Lloyd}(1996)}]{lloyd96}%
  \BibitemOpen
  \bibfield  {author} {\bibinfo {author} {\bibfnamefont {S.}~\bibnamefont
  {Lloyd}},\ }\href@noop {} {\bibfield  {journal} {\bibinfo  {journal}
  {Science}\ }\textbf {\bibinfo {volume} {273}},\ \bibinfo {pages} {1073}
  (\bibinfo {year} {1996})}\BibitemShut {NoStop}%
\bibitem [{\citenamefont {Poulin}\ \emph {et~al.}(2014)\citenamefont {Poulin},
  \citenamefont {Hastings}, \citenamefont {Wecker}, \citenamefont {Wiebe},
  \citenamefont {Doherty},\ and\ \citenamefont {Troyer}}]{poulin14}%
  \BibitemOpen
  \bibfield  {author} {\bibinfo {author} {\bibfnamefont {D.}~\bibnamefont
  {Poulin}}, \bibinfo {author} {\bibfnamefont {M.~B.}\ \bibnamefont
  {Hastings}}, \bibinfo {author} {\bibfnamefont {D.}~\bibnamefont {Wecker}},
  \bibinfo {author} {\bibfnamefont {N.}~\bibnamefont {Wiebe}}, \bibinfo
  {author} {\bibfnamefont {A.~C.}\ \bibnamefont {Doherty}}, \ and\ \bibinfo
  {author} {\bibfnamefont {M.}~\bibnamefont {Troyer}},\ }\href@noop {}
  {\bibfield  {journal} {\bibinfo  {journal} {arXiv preprint arXiv:1406.4920}\
  } (\bibinfo {year} {2014})}\BibitemShut {NoStop}%
\bibitem [{\citenamefont {Trout}\ and\ \citenamefont {Brown}(2015)}]{trout15}%
  \BibitemOpen
  \bibfield  {author} {\bibinfo {author} {\bibfnamefont {C.~J.}\ \bibnamefont
  {Trout}}\ and\ \bibinfo {author} {\bibfnamefont {K.~R.}\ \bibnamefont
  {Brown}},\ }\href@noop {} {\bibfield  {journal} {\bibinfo  {journal}
  {International Journal of Quantum Chemistry}\ }\textbf {\bibinfo {volume}
  {115}},\ \bibinfo {pages} {1296} (\bibinfo {year} {2015})}\BibitemShut
  {NoStop}%
\bibitem [{\citenamefont {Wecker}\ \emph {et~al.}(2015)\citenamefont {Wecker},
  \citenamefont {Hastings}, \citenamefont {Wiebe}, \citenamefont {Clark},
  \citenamefont {Nayak},\ and\ \citenamefont {Troyer}}]{wecker15}%
  \BibitemOpen
  \bibfield  {author} {\bibinfo {author} {\bibfnamefont {D.}~\bibnamefont
  {Wecker}}, \bibinfo {author} {\bibfnamefont {M.~B.}\ \bibnamefont
  {Hastings}}, \bibinfo {author} {\bibfnamefont {N.}~\bibnamefont {Wiebe}},
  \bibinfo {author} {\bibfnamefont {B.~K.}\ \bibnamefont {Clark}}, \bibinfo
  {author} {\bibfnamefont {C.}~\bibnamefont {Nayak}}, \ and\ \bibinfo {author}
  {\bibfnamefont {M.}~\bibnamefont {Troyer}},\ }\href@noop {} {\bibfield
  {journal} {\bibinfo  {journal} {Physical Review A}\ }\textbf {\bibinfo
  {volume} {92}},\ \bibinfo {pages} {062318} (\bibinfo {year}
  {2015})}\BibitemShut {NoStop}%
\bibitem [{\citenamefont {Hastings}\ \emph {et~al.}(2015)\citenamefont
  {Hastings}, \citenamefont {Wecker}, \citenamefont {Bauer},\ and\
  \citenamefont {Troyer}}]{hastings15}%
  \BibitemOpen
  \bibfield  {author} {\bibinfo {author} {\bibfnamefont {M.~B.}\ \bibnamefont
  {Hastings}}, \bibinfo {author} {\bibfnamefont {D.}~\bibnamefont {Wecker}},
  \bibinfo {author} {\bibfnamefont {B.}~\bibnamefont {Bauer}}, \ and\ \bibinfo
  {author} {\bibfnamefont {M.}~\bibnamefont {Troyer}},\ }\href@noop {}
  {\bibfield  {journal} {\bibinfo  {journal} {Quantum Information \&
  Computation}\ }\textbf {\bibinfo {volume} {15}},\ \bibinfo {pages} {1}
  (\bibinfo {year} {2015})}\BibitemShut {NoStop}%
\bibitem [{\citenamefont {Jones}(2013{\natexlab{b}})}]{jones13b}%
  \BibitemOpen
  \bibfield  {author} {\bibinfo {author} {\bibfnamefont {C.}~\bibnamefont
  {Jones}},\ }\href@noop {} {\bibfield  {journal} {\bibinfo  {journal}
  {Physical Review A}\ }\textbf {\bibinfo {volume} {87}},\ \bibinfo {pages}
  {022328} (\bibinfo {year} {2013}{\natexlab{b}})}\BibitemShut {NoStop}%
\bibitem [{\citenamefont {Eastin}(2013)}]{eastin13}%
  \BibitemOpen
  \bibfield  {author} {\bibinfo {author} {\bibfnamefont {B.}~\bibnamefont
  {Eastin}},\ }\href@noop {} {\bibfield  {journal} {\bibinfo  {journal}
  {Physical Review A}\ }\textbf {\bibinfo {volume} {87}},\ \bibinfo {pages}
  {032321} (\bibinfo {year} {2013})}\BibitemShut {NoStop}%
\bibitem [{\citenamefont {Paetznick}\ and\ \citenamefont
  {Reichardt}(2013)}]{Paetznick13}%
  \BibitemOpen
  \bibfield  {author} {\bibinfo {author} {\bibfnamefont {A.}~\bibnamefont
  {Paetznick}}\ and\ \bibinfo {author} {\bibfnamefont {B.~W.}\ \bibnamefont
  {Reichardt}},\ }\href {\doibase 10.1103/PhysRevLett.111.090505} {\bibfield
  {journal} {\bibinfo  {journal} {Phys. Rev. Lett.}\ }\textbf {\bibinfo
  {volume} {111}},\ \bibinfo {pages} {090505} (\bibinfo {year}
  {2013})}\BibitemShut {NoStop}%
\bibitem [{\citenamefont {Duclos-Cianci}\ and\ \citenamefont
  {Poulin}(2015)}]{duclos15}%
  \BibitemOpen
  \bibfield  {author} {\bibinfo {author} {\bibfnamefont {G.}~\bibnamefont
  {Duclos-Cianci}}\ and\ \bibinfo {author} {\bibfnamefont {D.}~\bibnamefont
  {Poulin}},\ }\href {\doibase 10.1103/PhysRevA.91.042315} {\bibfield
  {journal} {\bibinfo  {journal} {Phys. Rev. A}\ }\textbf {\bibinfo {volume}
  {91}},\ \bibinfo {pages} {042315} (\bibinfo {year} {2015})}\BibitemShut
  {NoStop}%
\bibitem [{\citenamefont {Gottesman}\ and\ \citenamefont
  {Chuang}(1999{\natexlab{b}})}]{GC01a}%
  \BibitemOpen
  \bibfield  {author} {\bibinfo {author} {\bibfnamefont {D.}~\bibnamefont
  {Gottesman}}\ and\ \bibinfo {author} {\bibfnamefont {I.}~\bibnamefont
  {Chuang}},\ }\href@noop {} {\bibfield  {journal} {\bibinfo  {journal}
  {Nature}\ }\textbf {\bibinfo {volume} {402}},\ \bibinfo {pages} {390}
  (\bibinfo {year} {1999}{\natexlab{b}})}\BibitemShut {NoStop}%
\bibitem [{\citenamefont {Bocharov}\ \emph
  {et~al.}(2015{\natexlab{a}})\citenamefont {Bocharov}, \citenamefont
  {Roetteler},\ and\ \citenamefont {Svore}}]{bocharov15}%
  \BibitemOpen
  \bibfield  {author} {\bibinfo {author} {\bibfnamefont {A.}~\bibnamefont
  {Bocharov}}, \bibinfo {author} {\bibfnamefont {M.}~\bibnamefont {Roetteler}},
  \ and\ \bibinfo {author} {\bibfnamefont {K.~M.}\ \bibnamefont {Svore}},\
  }\href@noop {} {\bibfield  {journal} {\bibinfo  {journal} {Physical Review
  A}\ }\textbf {\bibinfo {volume} {91}},\ \bibinfo {pages} {052317} (\bibinfo
  {year} {2015}{\natexlab{a}})}\BibitemShut {NoStop}%
\bibitem [{\citenamefont {Bocharov}\ \emph
  {et~al.}(2015{\natexlab{b}})\citenamefont {Bocharov}, \citenamefont
  {Roetteler},\ and\ \citenamefont {Svore}}]{bocharov15b}%
  \BibitemOpen
  \bibfield  {author} {\bibinfo {author} {\bibfnamefont {A.}~\bibnamefont
  {Bocharov}}, \bibinfo {author} {\bibfnamefont {M.}~\bibnamefont {Roetteler}},
  \ and\ \bibinfo {author} {\bibfnamefont {K.~M.}\ \bibnamefont {Svore}},\
  }\href {\doibase 10.1103/PhysRevLett.114.080502} {\bibfield  {journal}
  {\bibinfo  {journal} {Phys. Rev. Lett.}\ }\textbf {\bibinfo {volume} {114}},\
  \bibinfo {pages} {080502} (\bibinfo {year} {2015}{\natexlab{b}})}\BibitemShut
  {NoStop}%
\bibitem [{\citenamefont {Li}(2015)}]{ying15}%
  \BibitemOpen
  \bibfield  {author} {\bibinfo {author} {\bibfnamefont {Y.}~\bibnamefont
  {Li}},\ }\href@noop {} {\bibfield  {journal} {\bibinfo  {journal} {New
  Journal of Physics}\ }\textbf {\bibinfo {volume} {17}},\ \bibinfo {pages}
  {023037} (\bibinfo {year} {2015})}\BibitemShut {NoStop}%
\bibitem [{Note1()}]{Note1}%
  \BibitemOpen
  \bibinfo {note} {The Gridsynth package is available at
  http://www.mathstat.dal.ca/~selinger/newsynth/}\BibitemShut {NoStop}%
\bibitem [{\citenamefont {Duclos-Cianci}()}]{duclosPC}%
  \BibitemOpen
  \bibfield  {author} {\bibinfo {author} {\bibnamefont {Duclos-Cianci}},\
  }\href@noop {} {}\bibinfo {howpublished} {private communication}\BibitemShut
  {NoStop}%
\bibitem [{\citenamefont {Knill}(2005)}]{Knill05}%
  \BibitemOpen
  \bibfield  {author} {\bibinfo {author} {\bibfnamefont {E.}~\bibnamefont
  {Knill}},\ }\href@noop {} {\bibfield  {journal} {\bibinfo  {journal}
  {Nature}\ }\textbf {\bibinfo {volume} {434}},\ \bibinfo {pages} {39}
  (\bibinfo {year} {2005})}\BibitemShut {NoStop}%
\bibitem [{\citenamefont {Campbell}\ and\ \citenamefont
  {Howard}(2016{\natexlab{a}})}]{campbell16}%
  \BibitemOpen
  \bibfield  {author} {\bibinfo {author} {\bibfnamefont {E.~T.}\ \bibnamefont
  {Campbell}}\ and\ \bibinfo {author} {\bibfnamefont {M.}~\bibnamefont
  {Howard}},\ }\href@noop {} {\bibfield  {journal} {\bibinfo  {journal} {arXiv
  preprint arXiv:1606.01906}\ } (\bibinfo {year}
  {2016}{\natexlab{a}})}\BibitemShut {NoStop}%
\bibitem [{\citenamefont {Campbell}\ and\ \citenamefont
  {Howard}(2016{\natexlab{b}})}]{campbell16b}%
  \BibitemOpen
  \bibfield  {author} {\bibinfo {author} {\bibfnamefont {E.~T.}\ \bibnamefont
  {Campbell}}\ and\ \bibinfo {author} {\bibfnamefont {M.}~\bibnamefont
  {Howard}},\ }\href@noop {} {\bibfield  {journal} {\bibinfo  {journal} {arXiv
  preprint arXiv:1606.01904}\ } (\bibinfo {year}
  {2016}{\natexlab{b}})}\BibitemShut {NoStop}%
\bibitem [{\citenamefont {Wang}\ \emph {et~al.}(2013)\citenamefont {Wang},
  \citenamefont {Berry}, \citenamefont {de~Oliveira},\ and\ \citenamefont
  {Sanders}}]{Wang13}%
  \BibitemOpen
  \bibfield  {author} {\bibinfo {author} {\bibfnamefont {D.-S.}\ \bibnamefont
  {Wang}}, \bibinfo {author} {\bibfnamefont {D.~W.}\ \bibnamefont {Berry}},
  \bibinfo {author} {\bibfnamefont {M.~C.}\ \bibnamefont {de~Oliveira}}, \ and\
  \bibinfo {author} {\bibfnamefont {B.~C.}\ \bibnamefont {Sanders}},\ }\href
  {\doibase 10.1103/PhysRevLett.111.130504} {\bibfield  {journal} {\bibinfo
  {journal} {Phys. Rev. Lett.}\ }\textbf {\bibinfo {volume} {111}},\ \bibinfo
  {pages} {130504} (\bibinfo {year} {2013})}\BibitemShut {NoStop}%
\bibitem [{\citenamefont {Wang}\ and\ \citenamefont {Sanders}(2015)}]{Wang15}%
  \BibitemOpen
  \bibfield  {author} {\bibinfo {author} {\bibfnamefont {D.-S.}\ \bibnamefont
  {Wang}}\ and\ \bibinfo {author} {\bibfnamefont {B.~C.}\ \bibnamefont
  {Sanders}},\ }\href {http://stacks.iop.org/1367-2630/17/i=4/a=043004}
  {\bibfield  {journal} {\bibinfo  {journal} {New Journal of Physics}\ }\textbf
  {\bibinfo {volume} {17}},\ \bibinfo {pages} {043004} (\bibinfo {year}
  {2015})}\BibitemShut {NoStop}%
\bibitem [{Note2()}]{Note2}%
  \BibitemOpen
  \bibinfo {note} {Private communication with Alex Bocharov}\BibitemShut
  {NoStop}%
\end{thebibliography}
\end{document}